\documentclass[twocolumn,dcolumn,superscriptaddress,showpacs,amsmath,amssymb,showkeys]{revtex4}
\usepackage{hyperref}
\usepackage[right, modulo, displaymath, mathlines]{lineno}
\usepackage{url}
\usepackage{amsfonts}
\usepackage{amssymb}
\usepackage{amsmath}
\usepackage{graphicx}
\usepackage{psfrag}
\usepackage{mathrsfs}
\usepackage{stmaryrd}
\usepackage{subfigure}
\usepackage{color}
\usepackage{amssymb,amsfonts,amsmath}

\def\Bgamma{\mbox{\boldmath$\gamma$}}

\def\Btheta{\mbox{\boldmath$\theta$}}

\def\Bkappa{\mbox{\boldmath$\kappa$}}

\def\bzero{\mbox{\boldmath$0$}}

\def\bB{\mbox{\boldmath$ B$}}

\def\bE{\mbox{\boldmath$ E$}}
\def\bF{\mbox{\boldmath$ F$}}

\def\bJ{\mbox{\boldmath$ J$}}

\def\bL{\mbox{\boldmath$ L$}}

\def\be{\mbox{\boldmath$ e$}}

\def\bu{\mbox{\boldmath$ u$}}

\begin{document}
\title{Mechano-chemical spinodal decomposition: A phenomenological theory of phase transformations in multi-component, crystalline solids}
\author{Shiva Rudraraju \thanks{Mechanical Engineering, University of Michigan}, Anton Van der Ven\thanks{Materials, University of California Santa Barbara} \& Krishna Garikipati\thanks{Mechanical Engineering, Mathematics, University of Michigan, \href{mailto:krishna@umich.edu}{\nolinkurl{krishna@umich.edu}}}}
%

\begin{abstract}{We present a phenomenological treatment of diffusion-driven martensitic phase transformations in multi-component crystalline solids that arise from non-convex free energies in mechanical and chemical variables. The treatment describes diffusional phase transformations that are accompanied by symmetry breaking structural changes of the crystal unit cell and reveals the importance of a \emph{mechano-chemical spinodal}, defined as the region in strain-composition space where the free energy density function is non-convex. The approach is relevant to phase transformations wherein the structural order parameters can be expressed as linear combinations of strains relative to a high-symmetry reference crystal. The governing equations describing mechano-chemical spinodal decomposition are variationally derived from a free energy density function that accounts for interfacial energy via gradients of the rapidly varying strain and composition fields. A robust computational framework for treating the coupled, higher order diffusion and nonlinear strain gradient elasticity problems is presented. Because the local strains in an inhomogeneous, transforming microstructure can be finite, the elasticity problem must account for geometric nonlinearity. An evaluation of available experimental phase diagrams and first-principles free energies suggests mechano-chemical spinodal decomposition should occur in metal hydrides such as ZrH$_{2-2c}$.  The rich physics that ensues is explored in several numerical examples in two and three dimensions and the relevance of the mechanism is discussed in the context of important electrode materials for Li-ion batteries and high temperature ceramics.}

\end{abstract}
\maketitle

\section{Introduction and Formulation}
Spinodal decomposition is a continuous phase transformation mechanism occuring throughout a solid that is far enough from equilibrium for its free energy density to lose convexity with respect to an internal degree of freedom. The latter could include the local composition as in classical spinodal decomposition described by Cahn and Hilliard \cite{CahnHilliard1958}, or a suitable non-conserved order parameter as in Allen and Cahn's theory for spinodal ordering \cite{AllenCahn1979}. A key requirement for continuous transformations is that order parameters can be formulated to uniquely describe continuous paths connecting the various phases of the transformation. These phases then correspond to local minima on a single, continuous free energy density surface in that order parameter space. For classical spinodal decomposition inside a miscibility gap, all phases have the same crystal structure and symmetry, and the order parameter is simply the local composition. The existence of a single, continuous free energy density surface for all phases participating in a transformation implies, by geometric necessity, the presence of domains in order-parameter space where the free energy density is non-convex. Reaching those domains through supersaturation (by externally varying temperature or composition) makes the solid susceptible to a generalized spinodal decomposition.

Many important multi-component solids undergo phase transformations that couple diffusional redistribution of their components with a structural change of the crystallographic unit cell. One prominent example is the decomposition that occurs when cubic yttria-stabilized zirconia Zr$_{1-c}$Y$_c$O$_{2-c/2}$ is quenched into a two-phase equilibrium region between tetragonal Zr$_{1-c}$Y$_c$O$_{2-c/2}$ having low Y composition and cubic Zr$_{1-c}$Y$_c$O$_{2-c/2}$ having a high Y composition. Another occurs in Li battery electrodes made of spinel Li$_{c}$Mn$_{2}$O$_{4}$. Discharging to low voltages causes the compound to transform from cubic LiMn$_{2}$O$_{4}$ to tetragonal Li$_{2}$Mn$_{2}$O$_{4}$ through a two-phase diffusional phase transformation mechanism. As with simple diffusional phase transformations, these coupled diffusional/martensitic phase transformations can either occur through a nucleation and growth mechanism, or, if certain symmetry requirements are met, through a continuous mechanism due to an onset of an instability with respect to composition and/or a structural order parameter. 

Here we present a treatment of coupled diffusional/martensitic phase transformations triggered by instabilities with respect to both strain and composition. These phase transformations are characterized by a \emph{mechano-chemical spinodal} that is defined as a non-convex region of the free energy density function in the strain-composition space. The possibility of a mechano-chemical spinodal decomposition is motivated by recent first-principles studies of martensitic transformations in transition metal hydrides where a high temperature cubic phase is predicted to display negative curvatures with respect to strain, thus making strain a natural order parameter to distinguish the cubic parent phase from its low temperature tetragonal daughter phases \cite{ThomasAVDV}. In addition, the coupling with composition degrees of freedom (e.g. through the introduction of hydrogen vacancies in the metal hydrides) allows for the possibility that the free energy also exhibits a negative curvature with respect to the composition. Mechano-chemical spinodal decomposition, therefore, is a phenomenon that is likely present in many multi-component materials but has to date been overlooked as a mechanism by which a high symmetry phase can decompose martensitically and through diffusional redistribution upon quenching into a two phase regime. Structural phase transformations in solids driven by instability with respect to an internal shuffle of the atoms within the unit cell have been treated rigorously in the literature with coupled Cahn-Hilliard and Allen-Cahn approaches [5, 6, 7, 8]; but mechano-chemical spinodal decomposition is fundamentally different and necessitates a coupled treatment of both the strain and composition instabilities.

Our treatment is based on a generalized, Landau-type free energy density function that couples strain and composition instabilities. The governing equations of mechano-chemical spinodal decomposition, obtained by variational principles, generalize the classical equations of the Cahn-Hilliard formulation \cite{CahnHilliard1958}, and of nonlinear gradient elasticity \cite{Toupin1962,Barsch1984} by coupling these systems. The ability to solve this complex, nonlinear, strain- and composition-gradient-driven, mechano-chemical system for sufficiently general initial and boundary value problems in two and three dimensions also has been lacking heretofore. We introduce the computational framework to obtain such solutions in general, three-dimensional solids. This newfound capability allows us to then reinforce our discussion of the phenomenology with dynamics predicted by the numerical solutions. 

\subsection{The mechano-chemical spinodal in two dimensions}
\label{sec:2dimMechanochem} 
\label{sec:introduction:mechanochemfree energy}
For accessibility of the arguments, we first consider the two-dimensional analogue of the cubic to tetragonal transformation: the square to rectangle transformation. The high-symmetry square lattice will serve as the reference crystal relative to which strains are measured. Lower symmetry lattices that can be derived from the square lattice by homogeneous strain include the rectangle, the diamond and lattices where there are no constraints on the cell lengths and their angles. 
\begin{figure}[!hbt]
\vspace{0.0cm}
  \centering
  \includegraphics[width=0.5\textwidth]{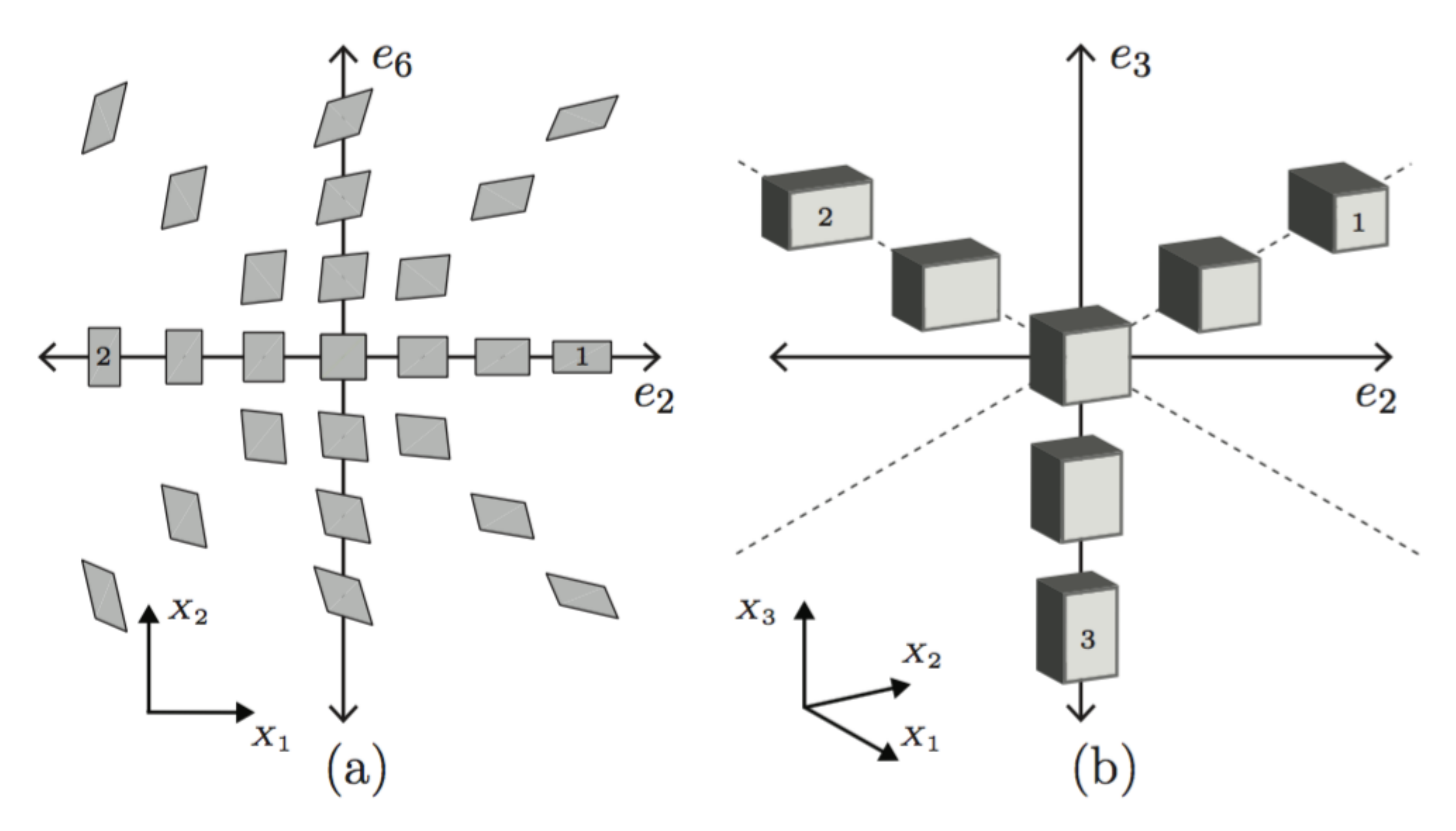}
\caption{(a) Square to rectangle transition (2D) in reparametrized strain space. (b) Cubic to tetragonal transition (3D) in reparametrized strain space  (Equation[\ref{strainmeasures3D}]). Lattice vectors are labelled by the corresponding coordinate directions {\tiny $X_1,X_2,X_3$} and the corresponding lower symmetry phases are labelled $1$,$~2$,$~3$. The deformations shown in the sub-figures are area/volume preserving, respectively. \label{fig:squareAndCubicDeformation}}
\end{figure}

We use the composition $c$, varying between $0$ and $1$, as our order parameter for the chemistry of our binary two-dimensional solid. Symmetry breaking structural changes are naturally described by the strain relative to the high symmetry square lattice. The elastic free energy density is also a function of strain. In both contexts, strain will in general be of finite magnitude. Following  Barsch \& Krumhansl \cite{Barsch1984}, we use the Green-Lagrange strain tensor, $\bE$, relative to the square lattice. Rotations are exactly neutralized in this strain measure; for any rigid body motion, $\bE = \bzero$ (Supporting Information). In two dimensions, the relevant strain components are 
$E_{11}, E_{22}$ and $E_{12} = E_{21}$. However, it is more convenient to use linear combinations of these components that transform according to the irreducible representations of the point group of the high symmetry square lattice. In two dimensions, these include $e_{1}=(E_{11}+E_{22})/\sqrt{2}$, $e_{2}=(E_{11}-E_{22})/\sqrt{2}$ and $e_{6}=\sqrt{2}E_{12}$. Here, $e_1$ and $e_6$ reduce to the dilatation and shear strain, respectively in the infinitesimal strain limit. The strain measure $e_2$ uniquely maps the square lattice into the two rectangular variants [Figure \ref{fig:squareAndCubicDeformation}a]: positive and negative $e_2$ generate the rectangles elongated along the global $X_1$ and $X_2$ directions, respectively. The equivalence of the rectangular variants under the point group symmetry of the square lattice ($e_2=0$) restricts the free energy density to even functions of $e_2$. 

\begin{figure}[!hbt]
\vspace{0.0cm}
  \centering
  \includegraphics[width=0.5\textwidth]{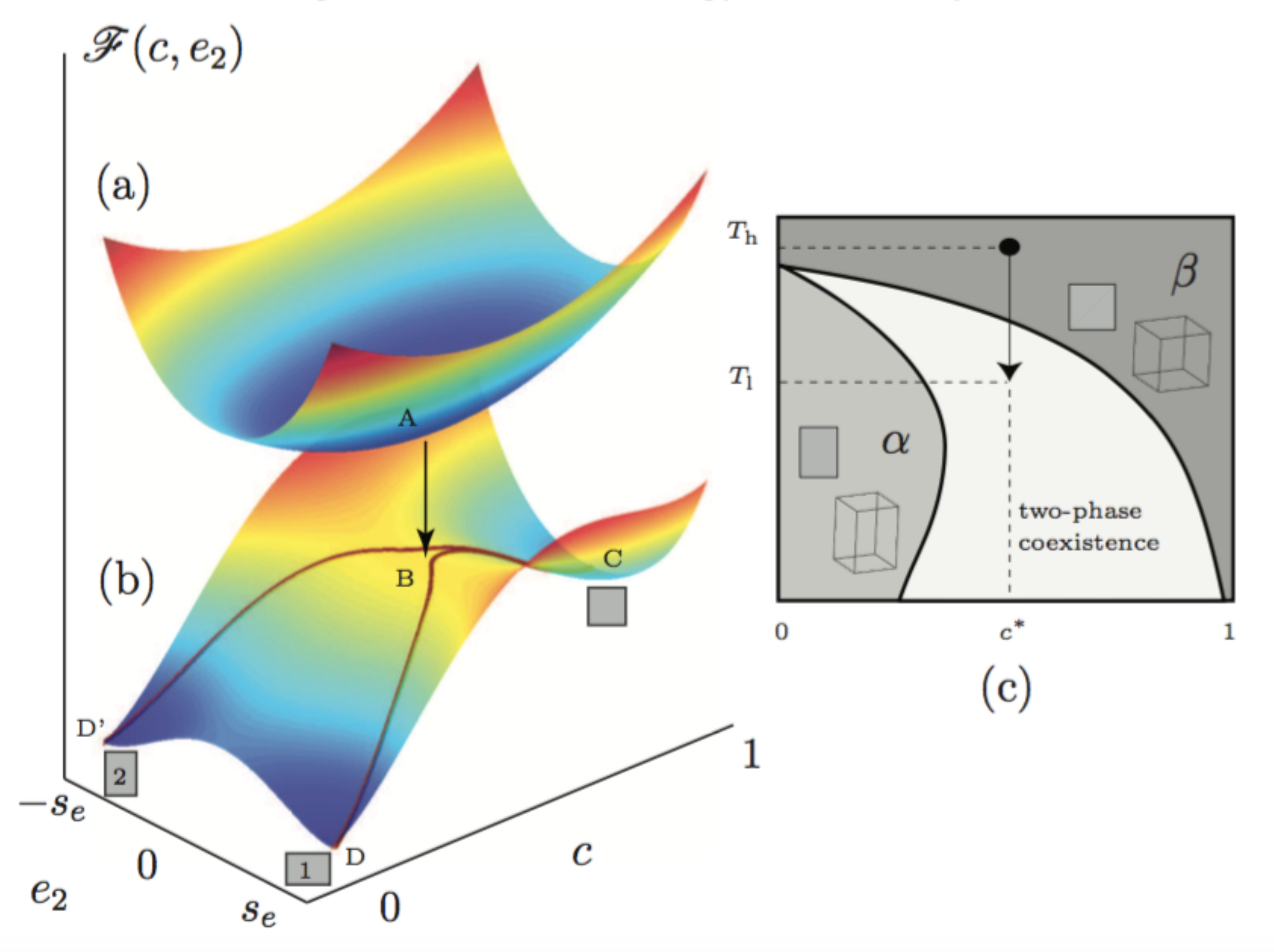}
\caption{(a) Free energy density for the 2D formulation at high temperature, and (b) at low temperature, showing the mechano-chemical spinodal along with two energy minimizing paths (brown lines) and their corresponding minimum energy strained structures. (c) Temperature-composition phase diagram. \label{fig:freeenergy2D}}
\end{figure}

If the crystalline solid has multiple chemical species, its free energy density dependence on $e_1$, $e_2$ and $e_6$ can change with composition, $c$. Figure~\ref{fig:freeenergy2D}a illustrates a free energy density surface, $\mathscr{F}(c,e_2)$, for a binary solid that, at higher temperature, forms a solid solution having square symmetry. In this case $\mathscr{F}$ is convex, a condition made precise by specifying positive eigenvalues of its Hessian matrix over the $\{c,e_2\}$ space. Additionally, $\mathscr{F}(c,0)$ is a minimum with respect to $e_2$ for fixed $c$, making the square phase stable for all $c$ at this temperature. However, at a lower temperature, $\mathscr{F}$ may lose convexity, inducing the notion of a \emph{mechano-chemical spinodal region}. We define it as the domain in $\{c,e_2\}$ space over which the Hessian matrix admits negative eigenvalues, as illustrated in Figure~\ref{fig:freeenergy2D}b. Here, we focus on the conditions $\partial^2 \mathscr{F}/\partial c^2 < 0$ and  $\partial^2 \mathscr{F}/\partial e_2^2 < 0$. The square phase ($e_2=0$) remains stable at high composition ($c \sim 1$) with positive eigenvalues of the Hessian (Figure~\ref{fig:freeenergy2D}b). But it is mechanically unstable at low composition, with $\partial^2 \mathscr{F}/\partial e_2^2 < 0$ for $(c,e_2) \sim (0,0)$, and has two symmetrically equivalent, stable, rectangular phases with $\partial^2 \mathscr{F}/\partial e_2^2 > 0$ at $(c,e_2)=(0,\pm s_\mathrm{e})$. While not shown in Figure~\ref{fig:freeenergy2D}b we assume that $\mathscr{F}$ is convex with respect to $e_1$ and $e_6$. Figure~\ref{fig:freeenergy2D}c illustrates a schematic temperature-composition phase diagram consistent with the free energy densities of Figure~\ref{fig:freeenergy2D}a-b. The square/cubic phase $\beta$, forms at high composition or at high temperature; the rectangle/tetragonal phase $\alpha$, forms at low composition and low temperature. A large two-phase region separates them. 

Zuzek et al. \cite{Zuzeketal1990} have obtained phase diagrams experimentally that are topologically equivalent to Figure \ref{fig:freeenergy2D}c for the $\mbox{ZrH}_{2-2c}$ system. Recent first principles calculations \cite{ThomasAVDV} have demonstrated the existence of a mechanical instability that exists in this system at low $c$ via non-convexity with respect to strain. Furthermore, the two-phase coexistence at low temperature also implies non-convexity with respect to composition as  we have demonstrated in supporting information. Figures \ref{fig:freeenergy2D}a and b therefore represent a two-dimensional analogue of the free energy for such systems.

Consider our model binary solid, annealed at high temperature $T_\mathrm{h}$, to form a solid solution in the square phase, $\beta$. Its state is at point A in Figure~\ref{fig:freeenergy2D}a, with $e_2 = 0$. It is then quenched into the two-phase region (Figure \ref{fig:freeenergy2D}c) with free energy density at point B in Figure~\ref{fig:freeenergy2D}b. For a quench at sufficiently high rate, the dimensions of the square lattice, controlled by strains $e_1$, $e_2$ and $e_6$, and the composition remain momentarily unchanged. However, since the state at point B satisfies $\partial^2 \mathscr{F}/\partial e^2_2 < 0$ and $\partial^2 \mathscr{F}/\partial c^2 < 0$, there exist thermodynamic driving forces for segregation by strain and composition within the mechano-chemical spinodal.  

Diffusion being substantially slower than elastic relaxation, the solid immediately deforms to either positive or negative $e_2$, driven towards a local minimum at constant $c$. These deformations due to the mechanical instability will happen like many martensitic transformations where a mix of symmetrically equivalent rectangular variants coexist to minimize macroscopic strain energy. For finite but moderate strain, the transformation could proceed coherently, even if the two symmetrically equivalent rectangular variants coexist. We neglect non-essential complexities of this process and assume that, instantly upon quenching, finitely sized neighborhoods of the solid deform homogeneously into one of these rectangular variants at the original composition. 

The solid also becomes susceptible to uphill diffusion because $\partial^2 \mathscr{F}/\partial c^2 < 0$ implies a negative diffusion coefficient. However, it does not occur at constant $e_2$, since the valleys traversing the local minima, $\partial\mathscr{F}/\partial e_2 = 0$, between the square lattice at $c = 1$ and the rectangular lattices at $c = 0$ span intervals of negative and positive $e_2$. Mechano-chemical spinodal decomposition sets in. Composition modulations are amplified: high $c$ regions strive to be more square (point C) while low $c$ regions strive to be more rectangular (points D or D$^\prime$). However, since coherency is maintained, some neighborhoods in the solid will be frustrated from attaining strains that ensure minima, $\partial\mathscr{F}/\partial e_2 = 0$, for local values of $c$. Coherency strain-induced free energy penalties arise to alter the driving forces for purely chemical spinodal decomposition. Movie S5 in the supporting information shows the evolution of the state $(c,e_2)$ of the material points on the free energy manifold $\mathscr{F}$.

\subsection{Mathematical formulation: three dimensions}
\label{sec:math}
Armed with the insight conveyed by the two-dimensional study, we next lay out the three-dimensional treatment, in which setting the considered phase transformations proceed via lattice deformation and diffusion. The arguments are made more concrete by considering the general mathematical form of the free energy density.

\subsubsection{Strain order parameters}
\label{sec:strain}

The strain measures $e_{1}$-$e_{6}$ are first redefined using the full Green-Lagrange strain tensor components in three dimensions:

\begin{align}
 & e_{1}=\frac{1}{\sqrt{3}}(E_{11}+E_{22}+E_{33}), \quad e_{2}=\frac{1}{\sqrt{2}}(E_{11}-E_{22}),\nonumber \\
 &  e_{3}=\frac{1}{\sqrt{6}}(E_{11}+E_{22}-2E_{33}), \quad e_{4}=\sqrt{2}E_{23}=\sqrt{2}E_{32},\nonumber \\
 &  e_{5}=\sqrt{2}E_{13}=\sqrt{2}E_{31}, \quad e_{6}=\sqrt{2}E_{12}=\sqrt{2}E_{21}
\label{strainmeasures3D}
\end{align}

In the limit of infinitesimal strains, $e_{1}$ describes the dilatation, while $e_4,\, e_5$ and $e_{6}$ reduce to shears. The point group operations of the cubic crystal leave $e_1$ invariant since it is the trace of $\bE$, while collecting each of the subsets $\{e_{2}, e_{3}\}$ and $\{e_4,\, e_5, e_{6}\}$ into a symmetry-invariant subspace whose elements transform into each other. The measures $e_{2}$ and $e_{3}$ are especially suited as order parameters to describe cubic to tetragonal distortions. All three tetragonal variants that emerge from the cubic reference crystal can be uniquely represented by these two measures. See Figure \ref{fig:squareAndCubicDeformation}b where tetragonal distortions of the cubic crystal along the $X_1$, $X_2$ and $X_3$ axes have been labelled as 1, 2 and 3, respectively. Deviations from the dashed lines in the $e_{2}$-$e_{3}$ space of Figure \ref{fig:squareAndCubicDeformation}b correspond to orthorhombic distortions of the cubic reference crystal. This role of $e_2$ and $e_3$ as structural order parameters to denote the degree of tetragonality, and to distinguish between the three tetragonal variants, complements their fundamental purpose as arguments of the elastic free energy density.
 
\begin{figure}[!hbt]
  \centering
  \includegraphics[width=0.5\textwidth]{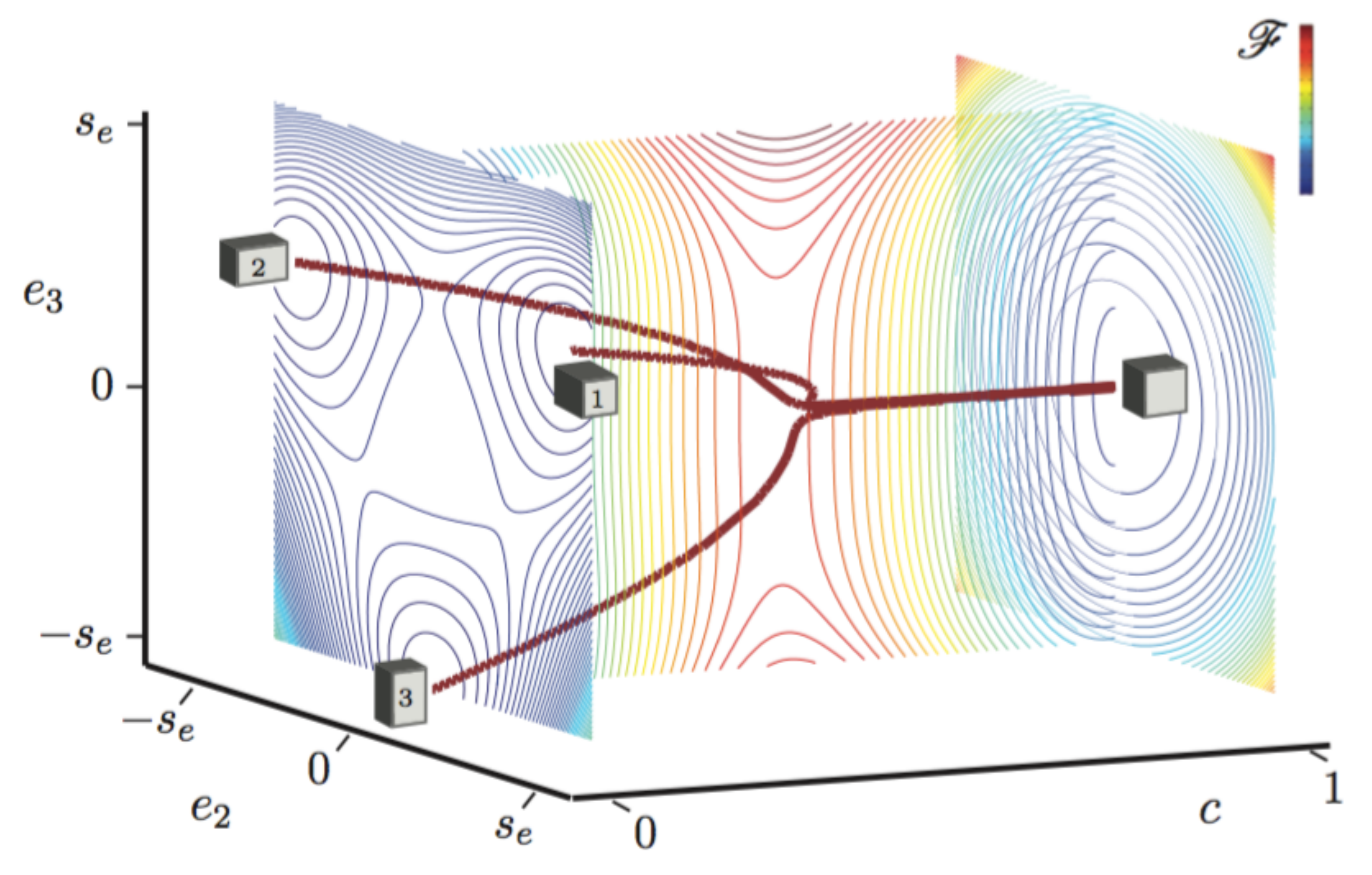}
\caption{Mechano-chemical spinodal for the 3D formulation depicted by contours of the free energy manifold along the $e_2$-$e_3$ and $e_3$-$c$ planes. The three energy minimizing paths (brown lines) and their corresponding energy minimizing strained structures and are also shown. \label{fig:freeenergy3D}}
\end{figure}

\subsubsection{The mechano-chemical spinodal in three dimensions} 

For brevity we write the strains as $\be = \{e_1,\dots,e_6\}$. We introduce further phenomenology by specifying that $\mathscr{F}(c, \be)$ is only one part of the total free energy density. It is a \emph{homogeneous} contribution whose composition and strain dependence cannot be additively separated.  Figure \ref{fig:freeenergy3D}, for example, illustrates contour plots of $\mathscr{F}(c,\be)$ on the $e_{2}$-$e_{3}$ and $e_{3}$-$c$ planes for a binary solid having a temperature-composition phase diagram similar to that of Figure~\ref{fig:freeenergy2D}c. To generate the tetragonal $\alpha$ phase as illustrated in that diagram, the homogeneous free energy density as a function of strain must qualitatively follow the contours of the $e_{2}$-$e_{3}$ plane at $c = 0$ in Figure \ref{fig:freeenergy3D}. Here, the tetragonal variants are local minimizers of $\mathscr{F}$ in $e_{2}$-$e_{3}$ space, with equal minima. In turn, to obtain the cubic $\beta$ phase, $\mathscr{F}(c,\be)$ must follow the contours of the $e_{2}$-$e_{3}$ plane at $c = 1$. Thus, $\mathscr{F}$ changes smoothly from convex with respect to $e_2$ and $e_3$ on the $c = 1$ plane to non-convex at $c = 0$ to form the three variants of the tetragonal $\alpha$ phase. Importantly, the planes at $c = 0$  and $c = 1$ themselves must be minimum energy surfaces to represent the tetragonal $\alpha$ and cubic $\beta$ phases, respectively. Movie S8 in the supporting information shows the evolution of the state $(c,e_2,e_3)$ of the material points on the free energy manifold $\mathscr{F}$.\\

\noindent \emph{Gradient regularization of the free energy density:}
Following  van der Waals \cite {vanderWaals1893}, and Cahn \& Hilliard \cite{CahnHilliard1958} in their treatment of non-uniform composition fields, we can extend the \emph{total} free energy density beyond $\mathscr{F}$ by writing it as a Taylor series, retaining terms that depend on the composition gradient, $\nabla c$. We extend this gradient dependence to the strain measures $\nabla\be$, as did Barsch \& Krumhansl \cite{Barsch1984} following Toupin \cite{Toupin1962}. (Also see Karatha and co-workers \cite{Sethna1991,Sethna1995}, for treatments using infinitesimal and finite strain, respectively.) These gradient dependences appear in a non-uniform free energy density $\mathscr{G}(c,\be,\nabla c,\nabla\be)$. Frame invariance of $\mathscr{F}$ and $\mathscr{G}$ is guaranteed since the members of $\be$ are linear combinations of the tensor components of $\bE$. They also must be invariant under point group operations of the cubic reference crystal.

\subsubsection{Free energy functional}
\label{sec:freeenergynonuniformsolid}
The crystal occupies a reference (undeformed) configuration $\Omega$ with boundary $\Gamma$. The total free energy, $\Pi$, is the integral of the free energy density $\mathscr{F} + \mathscr{G}$ over the solid with boundary contributions included. Thus, $\Pi$ is a functional of the composition $c$ and the displacement vector field $\bu$, from which the strains are derived (see Supporting Information):

\begin{equation}
	\Pi[ c , \bu] = \int\limits_{\Omega}  \left( \mathscr{F} +\mathscr{G} \right) \mathrm{d}V-\sum\limits_{i=1}^3\int\limits_{\Gamma_{T_i}}u_iT_i\mathrm{d}S.
\label{totalfreeenergy}
\end{equation}

\noindent where traction vector component $T_i$ is specified on the boundary subset $\Gamma_{T_i} \subset \Gamma$. Following the above authors we include only quadratic terms in the gradients, but in a generalization we also allow cross terms between $\nabla c$ and $\nabla \be$ in the non-uniform contribution $\mathscr{G}$, which can therefore be written as

\begin{align}
	\mathscr{G}( c , \be, \nabla c, \nabla\be) = &\frac{1}{2}\nabla c\cdot\Bkappa(c,\be)\nabla c \nonumber\\
	&+ \sum\limits_{\alpha,\beta}\frac{1}{2}\nabla e_\alpha\cdot\Bgamma^{\alpha\beta}(c,\be)\nabla e_\beta\nonumber\\
	&+ \sum\limits_{\alpha}\nabla c\cdot\Btheta^\alpha(c,\be)\nabla e_\alpha.
\label{compositiongradientenergyterm}
\end{align}

\noindent Here $\Bkappa$ is a symmetric tensor of composition gradient energy coefficients, each $\Bgamma^{\alpha\beta}$ ({\small $\alpha, \beta = 1,\dots,6$}) is a tensor of strain gradient energy coefficients, and each $\Btheta^\alpha$ is a tensor of the mixed, composition-strain gradient energy coefficients. Note that, in general,  these coefficients will be functions of the local composition and strain. The point group symmetry of the cubic reference crystal imposes constraints on the tensor components of $\Bkappa, \Bgamma^{\alpha\beta}$ and $\Btheta^\alpha$ as well as on the form of $\mathscr{F}$. 

While the gradient energies bestow greater accuracy upon the free energy description of solids with non-uniform composition and strain fields, they are essential at a more fundamental level if the homogeneous free energy density is non-convex. At compositions that render $\mathscr{F}$ non-convex, the absence of a gradient energy term will allow spinodal decomposition characterized by composition fluctuations of arbitrary fineness, thus leading to non-unique microstructures---a fundamentally unphysical result \cite{Hilliard}. With $\Bkappa \neq \bzero$, the composition gradient energy penalizes the interfaces wherein composition varies rapidly between high and low limits. This ensures physically realistic results, manifesting in unique microstructures with a mathematically well-posed formulation.

An essentially analogous situation exists with respect to the negative curvatures of $\mathscr{F}$ in the $e_2$-$e_3$ plane at low $c$, which drive the cubic lattice to distort into the tetragonal variants corresponding to the three free energy wells. Consider a solid with a homogeneous free energy density as in Figure \ref{fig:freeenergy3D} and a strain state lying between the valleys in the $e_2$-$e_3$ plane at $c = 0$. Absent the strain gradient energy, mechano-chemical spinodal decomposition would allow tetragonal variants of arbitrary fineness---an unphysical result, reflecting further mathematical ill-posedness. Retention of the strain gradient energy, ($\Bgamma^{\alpha\beta} \neq \bzero$) penalizes interfaces of sharply varying strain between tetragonal variants to ensure physically realistic results and unique microstructures from a mathematically well-posed formulation. This is well-understood in the literature that studies the formation of martensitic microstructures from non-convex free energy density functions\cite{Ball1987,Muller1999,Ball2011}.

\subsubsection{Governing equations of non-equilibrium chemistry}
\label{sec:formulation:chemicalDynamics} 
The free energy for non-homogeneous composition and strain fields, Eq. [\ref{totalfreeenergy}], must be a minimum at equilibrium. The state of a solid out of equilibrium will evolve to reduce the free energy $\Pi[ c , \bu]$.  In formulating a kinetic equation for the redistribution of atomic species through diffusion, we are guided by variational extremization of the free energy to identify the chemical potential, $\mu$. Details of this calculation appear as supporting information. The result follows:

\begin{align}
\mu = &\frac{\partial \mathscr{F}}{\partial c} -\nabla\cdot\left(\Bkappa\nabla c\right) + \nabla c\cdot\frac{\partial \Bkappa}{\partial c}\nabla c\nonumber\\
& + \sum\limits_{\alpha,\beta}\frac{1}{2}\nabla e_\alpha\cdot\frac{\partial\Bgamma^{\alpha\beta}}{\partial c}\nabla e_\beta\nonumber\\
&  + \sum\limits_{\alpha}\left(\nabla c\cdot\frac{\partial\Btheta^\alpha}{\partial c}\nabla e_\alpha - \nabla \cdot\left(\Btheta^\alpha\nabla e_\alpha\right)\right). 
\label{chempot}
\end{align}

\noindent For solids where $c$ tracks the composition of an interstitial element within a chemically inert host, such as Li in Li$_{c}$Mn$_{2}$O$_{4}$, $\mu$ in Eq. [\ref{chempot}] corresponds to the chemical potential of the interstitial element. If $c$ tracks the composition of a substitutional species, such as in alloys or on sublattices of complex compounds (e.g. the cation sublattice of yttria stabilized zirconia), $\mu$ is equal to the chemical potential difference between the substitutional species. 

The common phenomenological relation for the flux is $\bJ = -\bL(c,\be)\nabla\mu$, where $\bL$ is the Onsager transport tensor (see de Groot \& Mazur \cite{degrootmazur1984}). For an interstitial species, $\bL$ is related to a mobility \cite{avdvAccounts2013}, while it is a kinetic, intermixing coefficient for a binary substitutional solid \cite{avdvProgMatSci2010}. Inserting the flux in a mass conservation equation yields the strong form of the governing partial differential equation (PDE) for time-dependent mass transport. It is of fourth order in space due to the composition gradient dependence of $\mu$ in Equation [\ref{chempot}]. See Supporting Information for strong and weak forms of this PDE.

\subsubsection{Governing equations of mechanical equilibrium: Strain gradient elasticity}
\label{sec:formulation:mechanicalEquilibrium}

Mechanical equilibrium is assumed since elastic wave propagation typically is a much faster process than diffusional relaxation in crystalline solids. Equilibrium is imposed by extremizing the free energy functional with respect to the displacement field. Standard variational techniques lead to the weak and strong forms of strain gradient elasticity. The treatment is technical, for which reason we restrict ourselves to the constitutive relations that are counterparts to the chemical potential equation [\ref{chempot}] for chemistry. Coordinate notation is used for transparency of the tensor algebra, and summation is implied over repeated spatial index, $I$. Details appear in Supporting Information. The final form of the equations is complementary to Toupin's \cite{Toupin1962}, since our derivation is relative to the reference crystal, $\Omega$. 

With the deformation gradient $\bF$ being related to the Green-Lagrange strain as $E_{KL} = \frac{1}{2}(F_{iK}F_{iL} - \delta_{KL})$, the first Piola-Kirchhoff stress tensor and the higher-order stress tensor respectively, are given by

\begin{align}
P_{iJ}&= \sum\limits_{\alpha}\frac{\partial (\mathscr{F}+\mathscr{G})}{\partial e_\alpha}  \frac{\partial e_\alpha}{\partial F_{iJ}} +  \sum\limits_{\alpha}\frac{\partial \mathscr{G}}{\partial e_{\alpha,I}}\frac{\partial e_{\alpha,I}}{\partial F_{iJ}} \label{stressP} \\
B_{iJK} &= \sum\limits_{\alpha}\frac{\partial \mathscr{G}}{\partial e_{\alpha,I}}  \frac{\partial e_{\alpha,I}}{\partial F_{iJ,K}}\label{stressB}
\end{align}

The higher-order stress $\bB$, which is absent in classical, non-gradient elasticity \cite{TruesdellNoll} (and in earlier treatments of  mechano-chemistry \cite{Voorhees1,Larche}) makes the strong form of gradient elasticity a fourth order, nonlinear PDE in space (Supporting Information). The first three-dimensional solutions to general boundary value problems of Toupin's strain gradient elasticity  theory at finite strains were recently presented by the authors \cite{Rudrarajuetal2014}. 

%
%
\section{Results}
\label{sec:results}

\subsection{Two dimensional examples}
\label{sec:2Dnumegs}

We first consider a two-dimensional solid to better visualize the microstructures that can emerge from mechano-chemical spinodal decomposition. Plane strain elasticity is assumed, for which $E_{13},E_{23},E_{33} = 0$, giving $e_4, e_5 = 0$, and $e_3 = e_1/\sqrt{2}$, reducing Equations [\ref{strainmeasures3D}--\ref{stressB}] to two dimensions. The discussion on two-dimensional mechano-chemical spinodal decomposition (Figure \ref{fig:freeenergy2D}) holds: as a particular parameterization of $\mathscr{F}(c,e_1,e_2,e_6)$ we consider a regular solution model as a function of $c$ at zero strain. At $c=0$,  $\mathscr{F}(c,e_1,e_2,e_6)$ is double-welled in $e_2$ corresponding to the two rectangular variants. The gradient energy is $\mathscr{G}(\nabla c,\nabla e_{2})$ with a constant, isotropic $\Bkappa \neq \bzero$, and constant $\gamma^{22} \neq 0$, all other gradient coefficients being zero. While the fullest possible complexity of the coupling is not revealed by these simplifications, the aim here is to present the essential physics that is universal to mechano-chemical spinodal decomposition, postponing details of the more complex couplings and tensorial forms to future communications, where they will be derived for specific materials systems. See Supporting Information for specific forms of $\mathscr{F}\left( c,e_1,e_2,e_6\right)$ and $\mathscr{G}(\nabla c,\nabla e_2)$.

\begin{figure}[!hbt]
\vspace{0.0cm}
  \centering
  \includegraphics[width=0.5\textwidth]{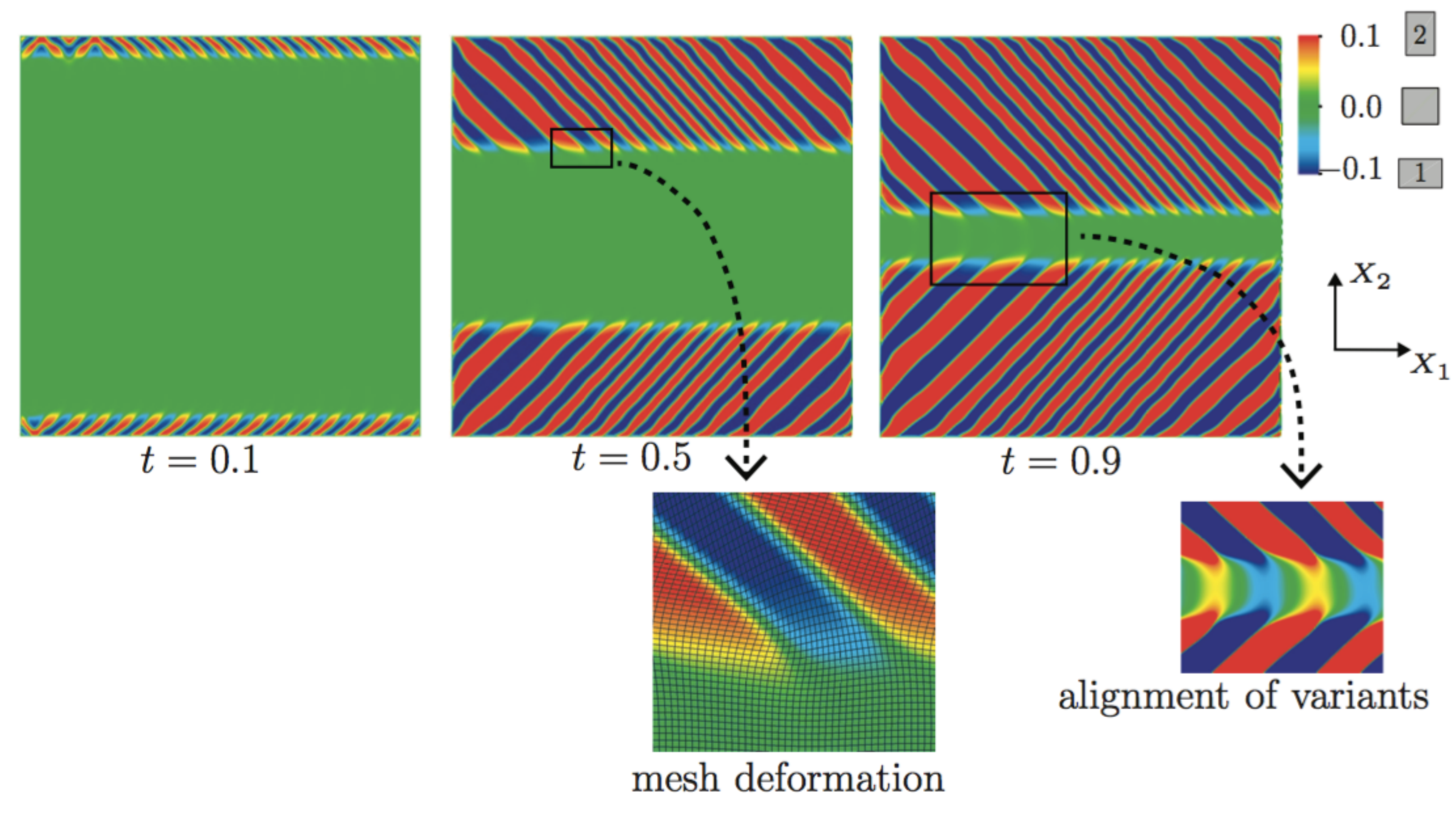}
\caption{Evolution of 2D microstructure during outflux from the top and bottom surfaces of a solid under plane strain. Contours show strain $e_2$. Note the legend and corresponding square/rectangular variant crystal structures. The deformation and accompanying twinned microstructure are seen clearly in the distorted mesh. \label{fig:timeEvolution}}
\end{figure}

Figure \ref{fig:timeEvolution} shows the evolution of microstructure over a $0.01 \times 0.01$ domain whose reference (initial) state has the square crystal structure and $c = 1$. The displacement component $u_1=10^{-5}$ on the right boundary ($X_1=0.01$), with the remaining boundaries fixed ($\bu = \bzero$). An outward flux is imposed on the top and bottom boundaries causing a decrease in composition, $c$, starting from the boundaries. As the composition falls, the homogeneous free energy density $\mathscr{F}$ loses convexity and the state of the material $(c,e_2)$ enters the mechano-chemical spinodal along $e_2 = 0$ (Figure \ref{fig:freeenergy2D}b). The continuing outward flux first drives material near the top and bottom boundaries fully into the regime where the rectangular crystal structure is stable. As explained for Figure \ref{fig:freeenergy2D}b, the negative curvature $\partial^2\mathscr{F}/\partial e_2^2 < 0$ creates thermodynamic driving forces that distort the square structure, shown in green, into rectangular variants, which form as red/blue laminae (all $e_2$ values). The coexistence of the parent, square structure with the daughter, rectangular variants also can give rise to cross-hatched microstructures that parallel the tweed microstructures described in the work of Karatha and co-workers \cite{Sethna1991,Sethna1995}. Movie S9 in the supporting information shows the formation of such a microstructure.

A laminar, micro-twinned microstructure develops as the two rectangular variants form, distinguished by the sign of $e_2$ (see legend). Note that $e_2$ remains continuous because of the penalization of strain gradients $\nabla e_2$, although discontinuities can develop in $\nabla e_2$, itself. Because our numerical framework uses basis functions that are continuous up to their first derivatives (see Methods and Supporting Information), $\nabla e_2$ is indeed discontinuous in the computations. The lamination accommodates the strain difference between the two rectangular variants to minimize the free energy. If strains were infinitesimal ($\vert e_1 \vert, \vert e_2 \vert, \vert e_6 \vert \gtrsim 0$) $e_2$ would correspond to shear along directions that are rotated $\pi/4$ radians from the crystal axes. But the strains in these microstructures can be finite (reflected in the range $-0.1 \le e_2 \le 0.1$ in Figure \ref{fig:timeEvolution}) and involve large rotations to accommodate the rectangular variants; this necessitates a finite-strain formulation, as shown in recent studies by Finel et al \cite{AlphonseFinel} comparing infinitesimal-strain and finite-strain formulations for polytwinned microstructure evolution in martensitic alloys, and by Hildebrand \& Miehe \cite{Hildebrand2012}.

The micro-twinning and coherency strains are seen in the distorted mesh of discretization (inset of top center panel). The undeformed mesh cells are squares, and hence the numerical discretization strikingly delineates the kinematics of the cubic to rectangular transformation, and highlights the rectangular twins as well as the concentrated distortions along the twin boundaries. Movie S6 in the supporting information shows such mesh distortion and formation of the rectangular twins with a different form of the  function $\mathscr{F}$.

\begin{figure}[!hbt]
\vspace{0.0cm}
  \centering
  \includegraphics[width=0.5\textwidth]{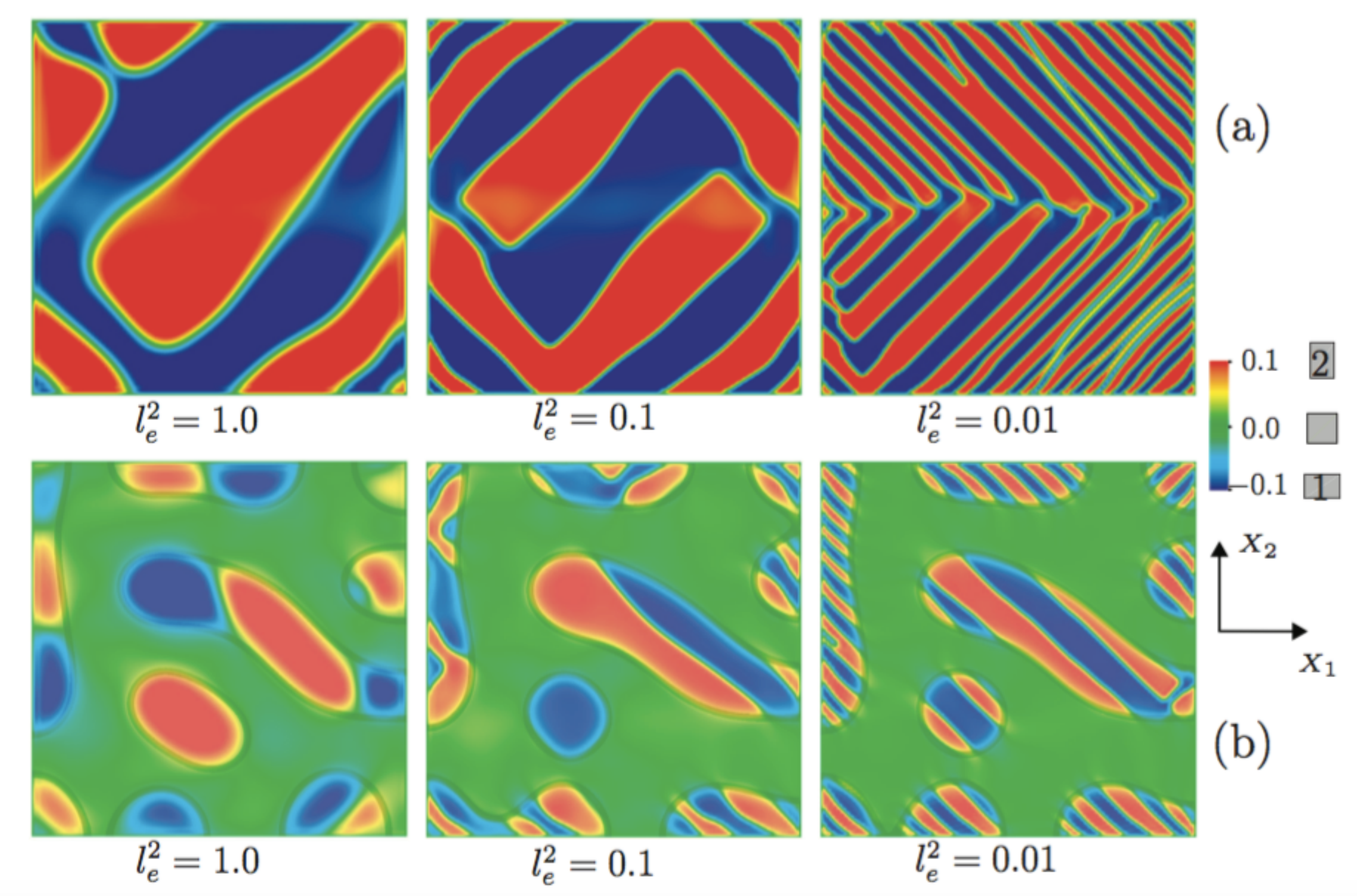}
\caption{Microstructure controlled by elastic gradient length scale parameter ($l_\mathrm{e}$). Shown are the final contours of $e_2$ for the simulation of (a) outflux from top and bottom surfaces, and (b) quenching. \label{fig:lengthScaleEffect}}
\end{figure}

In some cases, the long-range nature of elasticity forces like rectangular variants to align even when separated by an untransformed square phase. This is first seen in the finger-like extensions of strain contours from the rectangular variants into the square phase in Figure \ref{fig:timeEvolution}, followed by their alignment and eventual incorporation into laminae of the same variant (top right panel and its evolution shown as an inset). In this instance, although the micro-twins end at an invariant habit plane that is common with the untransformed square phase, the lattice parameters of the rectangular variants differ from those of the square phase, inducing elastic strain energy in the latter. The alignment lowers this strain energy, and also is seen in Figure \ref{fig:lengthScaleEffect}a. Note, however, that this is not a universal feature. Movie S2 shows instances where unlike variants come close to alignment.

The fineness of laminae depends on the strain gradient elasticity length scale $l_\mathrm{e} \sim \sqrt{\gamma^{22}/\sqrt{\sum_{\alpha,\beta=1,2,6}(\partial^2\mathscr{F}/\partial e_\alpha\partial e_\beta)^2}}$, as explored in Figure \ref{fig:lengthScaleEffect}. For Figure \ref{fig:lengthScaleEffect}a, the initial and boundary conditions are the same as in the example of Figure \ref{fig:timeEvolution}. Decreasing $l_\mathrm{e}$ weakens the penalty on the strain gradient $\nabla e_2$ across neighboring, unlike rectangular variants and allows more twin boundaries. Notably, self-similarity is not maintained between microstructures for different $l_\mathrm{e}$, even for the same initial and boundary value problem. We understand this to be the influence of elastic strain accommodation: To minimize the total free energy when the strain gradient penalty changes, the physics optimizes  twin boundaries via laminations of different sizes as well as different patterns. Importantly however, crystal symmetry admits non-vanishing strain gradient energy coefficients beyond $\gamma^{22}\neq 0$ used in these simulations (Supporting Information). Furthermore, the composition dependence of $\mathscr{F}$ could be more complex than the  simple regular solution model used here. Given the already strong effect of $l_\mathrm{e}$ alone, we conjecture that varying these forms will have a significant influence on the resulting coherent twin microstructure. The proper form, while guided by crystal symmetry arguments must ultimately be determined experimentally or by first-principles statistical mechanical methods.

The example in Figure \ref{fig:lengthScaleEffect}b further explores the influence of $l_\mathrm{e}$. In this suite of computations, the unstrained material with an initially square microstructure, convex free energy density and composition having random fluctuations about $c = 0.45$ was quenched to a low temperature and into the mechano-chemical spinodal. The local composition and strain evolve under the thermodynamic driving forces detailed in the context of Figure \ref{fig:freeenergy2D}. We draw attention, once again, to the changing identity of rectangular variants, their shapes and sizes depending on $l_\mathrm{e}$. Also note the progressively finer lamination of rectangular domains with decreasing $l_\mathrm{e}$. Such studies suggest how dynamic mechano-chemical spinodal decomposition can lead to an atlas of microstructures, which in turn will determine material properties. Movies S1-S4 in the supporting information show the evolution of some of these microstructures.

\subsection{A three dimensional study}
\label{sec:3Dnumegs}

This final example displays the full three-dimensional complexity of microstructures resulting from the mechano-chemical spinodal. We persist with the above simplifications aimed at presenting the essential physics of mechano-chemical spinodal decomposition, and postpone a full exploration of the coupling and anisotropies in coefficients to future work on specific materials systems. The free energy density function used for the three-dimensional study appears as supporting information. 

\begin{figure}[!hbt]
\vspace{0.0cm}
  \centering
  \includegraphics[width=0.5\textwidth]{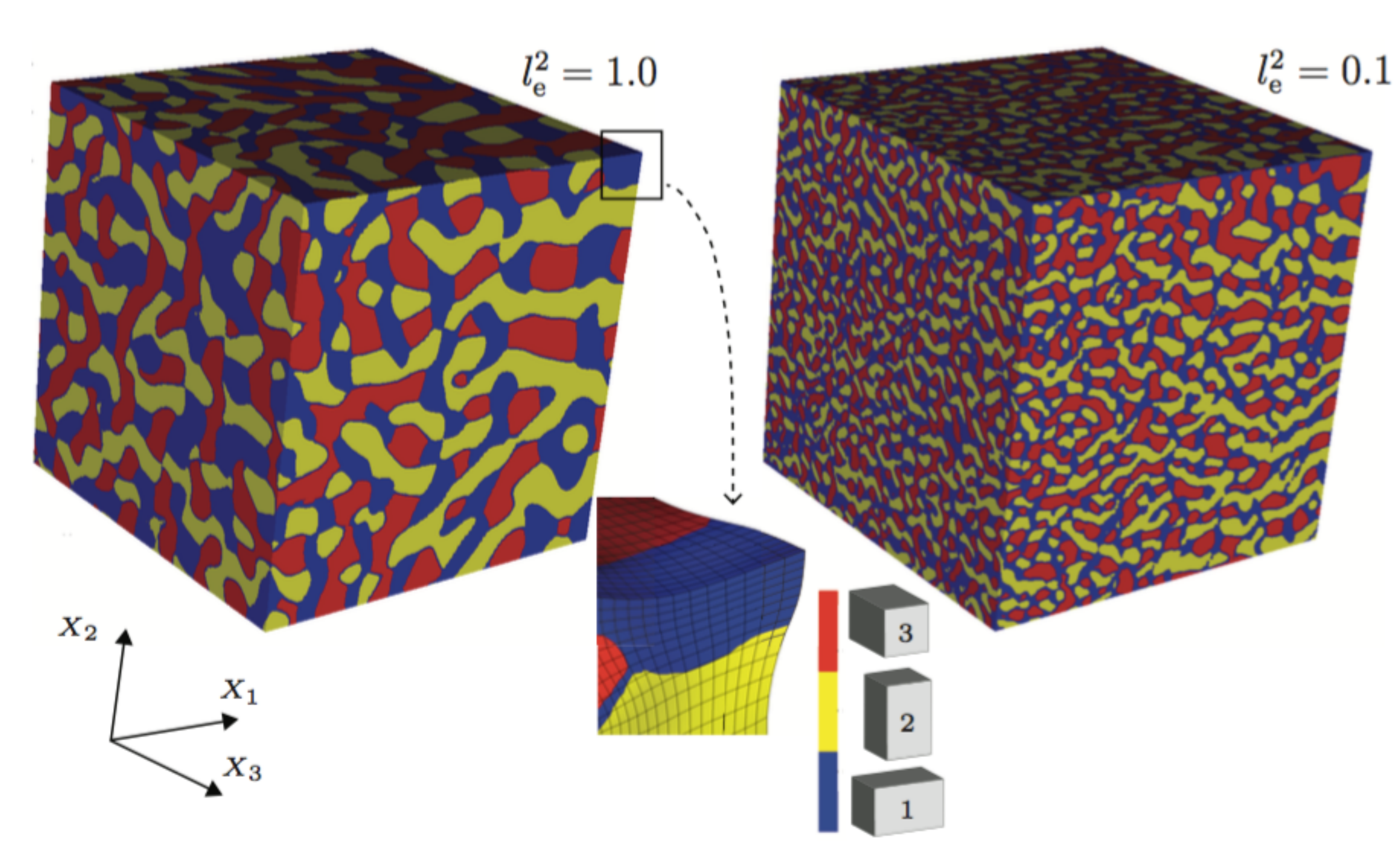}
\caption{Microstructure observed in 3D for different values of the elastic gradient length scale parameter ($l_\mathrm{e}$). The three tetragonal variants appear in blue (variant 1), yellow (variant 2) and red (variant 3) for $c=0$. The transformation strains are easily discerned in the distorted mesh. \label{fig:microStructure3D}}
\end{figure}

Figure \ref{fig:microStructure3D} shows the equilibrium microstructure that results in a solid that is initially in the cubic phase, $c = 1$, subject to an outflux on all surfaces. The domain is a unit cube with displacement components $u_2,u_3=0.01$ on the boundary $X_{1}=1$, with zero displacement, $\bu=\bzero$, on the boundary $X_{1}=0$. The cubic to tetragonal transformation takes place as $c \to 0$. Under these conditions all three tetragonal variants form, with an intricately interleaved microstructure for strain accommodation. Note the finer microstructures and changing pattern for smaller $l_\mathrm{e}$. The inset shows the surface straining around a corner, delineated by the distorted mesh lines. Observe the three, oriented, tetragonal variants formed by twinning deformation from the initial cubic structure. See Movie S7 in the supporting information for a detailed view of the three-dimensional structure of these individual tetragonal variants. Movie S10 in supporting information shows other three-dimensional microstructures whose cross-sectional planes bear closer resemblance to the plate-like structures predicted by two-dimensional computations. To the best of our knowledge, such computations of a cubic to tetragonal transformation with twinned variants whose microstructure is controlled by nonlinear, strain gradient interface energies, have not been previously presented.  

\section{Discussion}
\label{sec:conclusion}
Several kinetic mechanisms have been proposed to describe the decomposition of a solid upon entering a two-phase region \cite{Balluffi,PorterEasterling}. A commonly observed mechanism occurs through localized nucleation followed by diffusional growth that is mediated by the migration of interfaces separating the daughter phases from the parent phase. Nucleation and growth mechanisms are common when the phases participating in the transformation have little or no crystallographic relation to each other. The transformation then proceeds reconstructively, where the interphase interfaces are disordered or at best only semi-coherent. Numerical treatments of this mechanism rely on sharp-interface methods such as a level set approach \cite{Sethian}.

Other kinetic mechanisms of decomposition involving some degree of coherency are also possible when the participating phases share sufficient crystallographic commonality that order parameters can be defined describing continuous pathways connecting the parent and daughter phases. The structural changes of the crystal that accompany these transformations can fall in one of two categories. One subclass of structural transformations is driven by an internal shuffle, where the atomic arrangement within the unit cell of the crystal undergoes a symmetry breaking change. The unit cell vectors of the crytal may also undergo a change and lower the symmetry of the lattice, but this effect is secondary and in response to the atomic shuffle within the unit cell. The order parameters for the structural change therefore describe the atomic shuffles within the unit cell and are non conserved. The second subclass of structural transformations are driven by a symmetry reducing change of the lattice vectors of the crystal, with any atomic rearrangement of the basis of the crystallographic unit cell occurring in response to the symmetry breaking changes of the unit cell vectors. The natural order parameters describing these transitions are strains. The free energy in the first subclass will exhibit negative curvatures with respect to the non-conserved shuffle order parameters, while the free energy in the second subclass will have negative curvatures with respect to the relevant strain order parameters, as was recently predicted for the cubic to tetragonal transition of ZrH$_{2}$ \cite{ThomasAVDV}.

Decomposition reactions that require both diffusional redistribution and a structural change due to an {\it internal shuffle} have been treated successfully and rigorously with phase field approaches that combine Cahn-Hilliard with Allen-Cahn. The Cahn-Hilliard description accounts for the composition degrees of freedom while the Allen-Cahn component describes the evolution of the non-conserved shuffle order parameters required to distinguish the various phases of the transformation \cite{Wang1993,LQChen1995-1,LQChen1995-2,Hildebrand2012}. These treatments have included strain energy as a secondary effect serving only as a positive contribution to the overall free energy due to coherency strains. The approach is rigorous if the free energy density remains convex with respect to strain, with instabilities in the free energy appearing as a function of concentration and the non-conserved shuffle order parameters.

Here we have presented a mathematical formulation and an accompanying computational framework for decomposition reactions that combine diffusional redistribution with a structural change driven by a {\it symmetry breaking strain} of the crystallographic unit cell as opposed to an internal shuffle within the unit cell. Strains play a primary role, explicitly serving as order parameters to distinguish variants of a daughter phase that has a symmetry subgroup-group relationship with its parent phase due to a structural change of the crystallographic unit cell. Crucial to the description is that the driving force for the formation of the daughter from a supersaturated parent phase emerges from an instability with respect to composition. The treatment therefore combines Cahn-Hilliard for composition with a description of martensitic transformations at fixed concentration introduced by Barsch and Krumhansl \cite{Barsch1984}. The existence of simultaneous instabilities with respect to strain and composition has not been considered as a mechanism to describe decomposition reactions upon quenching into a two-phase region.

The possibility of such a mechano-chemical spinodal mechanism is motivated by recent first-principles studies of the cubic to tetragonal phase transformations of ZrH$_{2}$, where the free energy of the high temperature cubic form was predicted to become unstable with respect to strain upon cooling below the cubic to tetragonal second order transition temperature \cite{ThomasAVDV}. The ZrH$_{2-2c}$ hydride can accommodate large concentrations of hydrogen vacancies, $c$, and has a phase diagram that is topologically identical to that depicted in Figure 2c, with a two-phase region separating a hydrogen-rich tetragonal form of ZrH$_{2-2c_{\alpha}}$ from a cubic form of ZrH$_{2-2c_{\beta}}$ (with $c_{\alpha} < c_{\beta}$). See Zuzek et al. \cite{Zuzeketal1990}.\footnote{In their phase diagrams, Zuzek et al have an inverted composition axis relative to our notation. In their work, the tetragonal phase is labelled $\varepsilon$ and the cubic phase is $\delta$.} To be consistent with the predicted free energies for stoichiometric ZrH$_{2}$ (i.e. $c$=0) and the experimental T versus $c$ phase diagram with the form of Figure \ref{fig:freeenergy2D}c, the free energy of this hydride as a function of composition and strain (i.e. $e_{2}$ and $e_{3}$) should be similar to those depicted in Figures \ref{fig:freeenergy2D}a and b, and Figure \ref{fig:freeenergy3D}. Decomposition upon quenching cubic ZrH$_{2-2c}$ into the two-phase region can therefore proceed through a mechano-chemical spinodal decomposition mechanism.

Not only does the phenomenological description introduced in this work describe a new mechanism of decomposition that is qualitatively distinct from previous combined Cahn-Hilliard and Allen-Cahn approaches, its numerical solution also proves to be substantially more complex due to contributions from strain gradient terms. Indeed, even numerical solutions to general, three-dimensional boundary value problems of gradient elasticity at finite strain were not available until presented recently by the authors \cite{Rudrarajuetal2014}. Furthermore, as other authors have pointed out before, the use of strain metrics as order parameters makes a reliance on infinitesimal strains untenable due to the rigid rotations that accompany the finite strains characterizing most structural transformations \cite{Hildebrand2012,AlphonseFinel}. The use of finite strain metrics introduces geometric non-linearity into the problem, which while having been treated in past Cahn-Hilliard and Allen-Cahn approaches \cite{Hildebrand2012}, adds additional numerical challenges when also considering strain gradient contributions. These challenges have been overcome in this work as demonstrated by our three-dimensional numerical examples.

Mechano-chemical spinodal decomposition as described here can be expected in solids forming high temperature phases that exhibit dynamical instabilities at low temperature. An accumulating body of first-principles calculations of Born-Oppenheimer surfaces have shown that many high temperature phases are dynamically unstable at low temperature \cite{craievich1,Grimvall}, becoming stable at high temperature through large anharmonic vibrational excitations \cite{Zhong1,Finnis2,BhattacharyaAVDV,Souvatzis,ThomasAVDV}, usually through a second order phase transition. While such instabilities are frequently dominated by phonon modes describing an internal shuffle, a subset of chemistries become dynamically unstable at low temperature with respect to phonon modes that break the symmetry of the crystal unit cell \cite{ThomasAVDV, BhattacharyaAVDV}, an instability that can be described phenomenologically with strain order parameters. If, upon alloying such compounds, the high symmetry phase becomes stable by passing through a two-phase region, its free energy surface will resemble that of Figures~\ref{fig:freeenergy2D}b and \ref{fig:freeenergy3D}, and thereby make the solid susceptible to the mechano-chemical spinodal decomposition mechanism described here.

While first-principles and experimental evidence suggests mechano-chemical spinodal decomposition should occur in ZrH$_{2-2c}$, we expect it to occur in a wide range of other chemistries as well. One possible example, as described in the introduction, is the decomposition of cubic yttria stabilized zirconia \cite{Heuer, Levi} upon quenching from the high temperature cubic phase into a two-phase region separating a low-Y tetragonal phase from a Y-rich cubic phase. Past treatments of this transformation \cite{Wang1993,LQChen1995-1,LQChen1995-2} relied on non-conserved order parameters to distinguish the different tetragonal variants from each other and from the cubic parent phase. The elastic part of the free energy density function was parameterized using linearized elasticity and infinitesimal strains, as other authors have also done \cite{Bouville,GarthWells}. The chemical part of the free energy density was assumed to have a negative curvature as a function of the non-conserved order parameter at Zr-rich compositions, but made up of convex (and quadratic) potentials with respect to strain. We reiterate that such a treatment rests on the implicit assumption that internal shuffles drive the cubic to tetragonal transformation, while the free energy as a function of strain remains convex for all relevant values of the non-conserved order parameters. 

The possibility also exists that a coarse-grained free energy density of cubic $\mbox{ZrO}_2$ exhibits negative curvatures with respect to strain below the cubic to tetragonal transition temperature making the formalism developed here an accurate description of decomposition reactions of yttria stabilized zirconia. While the precise mechanism and nature of the cubic to tetragonal transition of pure $\mbox{ZrO}_2$ remains to be resolved \cite{Finnis2}, first-principles calculations predict that the cubic form of $\mbox{ZrO}_2$ is dynamically unstable with respect to transformation to the tetragonal variant \cite{Jomard}. A rigorous statistical mechanics treatment \cite{Zhong1,ThomasAVDV} is required to determine whether the cubic to tetragonal transition of pure $\mbox{ZrO}_2$ is accompanied by a change in the sign of the curvature of the free energy density with respect to $e_2$ and $e_3$. If this proves to be the case, yttria stabilized zirconia should also be susceptible to mechano-chemical spinodal decomposition upon quenching, consistent with the coherent spinodal microstructures between tetragonal and cubic phases observed in single crystal regions of quenched cubic $\mbox{Zr}_{1-c}\mbox{Y}_x\mbox{O}_{2-c/2}$  \cite{Levi}.

A mechano-chemical spinodal may also play a role in a variety of important electrode materials for Li-ion batteries, and intercalation compounds considered for two-dimensional nano-electronics. These include cubic $\mbox{LiMn}_{2}\mbox{O}_{4}$ transforming to tetragonal $\mbox{Li}_2\mbox{Mn}_{2}\mbox{O}_{4}$ \cite{Thackeray}. While most mechano-chemical spinodal transitions will occur in three dimensions, many of the qualitative features of these transitions are more conveniently illustrated in two dimensions. Our two-dimensional studies should also prove relevant to understanding mechano-chemical phase transformations in two-dimensional layered materials for nano electronics. Materials such as $\mbox{TaS}_2$, are susceptible to Peierls instabilities upon variation of the composition of adsorbed or intercalated guest species that donate to or extract electrons from the sheet-like host \cite{Rossnagel}. Our phenomenological treatment by introduction of the concept of a mechano-chemical spinodal, coupled with gradient stabilization of the ensuing non-convex free energy in strain-composition space, thus offers a framework with potential for extension to a wide range of phase transformation phenomena.

With regard to the fundamental thermodynamic underpinnings of analogous processes, the phenomenological model introduced here can also be used to describe temperature driven martensitic transformations. The composition, $c$, would then be analogous to the internal energy density, $\mathscr{U}$, while the chemical potential, $\mu$, would be analogous to the temperature, $T$. Due to the presence of temperature gradients throughout the solid, though, the starting point must be a formulation of the entropy density, $\mathscr{S}$, (as opposed to the Helmholtz free energy) as a functional of  spatially varying $\mathscr{U}$ and displacements, $\bu$. In analogy with Equation (\ref{totalfreeenergy}), the total entropy can be written as a volume integral over a homogeneous entropy density that depends on the local internal energy density and strain $\bar{\mathscr{S}}(\mathscr{U},\be)$, as well as a non-uniform entropy density contribution expressed in terms of gradients in internal energy density and strain ($\nabla \mathscr{U}$, $\nabla\be$). Variational maximization of the entropy will yield mechanical equilibrium equations as well as an expression for the temperature $T$ (strictly speaking, for its reciprocal, $1/T$), similar to Equation~(\ref{chempot}) for the chemical potential. Due to the presence of gradient terms, $\nabla\mathscr{U}$ and  $\nabla \be$, the temperature will not only be a function of the local internal energy density and strain, but also gradients in these field variables. As with the diffusion problem treated here, heat flow can be described with an Onsager expression relating the heat flux to a gradient in temperature. The possibility exists that the homogeneous entropy $\bar{\mathscr{S}}$ exhibits instabilities with respect to internal energy density $\mathscr{U}$, allowing for spinodal decomposition with respect to the redistribution of $\mathscr{U}$ in a manner that is similar to the well understood problem of spinodal decomposition with respect to composition. The treatment introduced here can therefore describe (upon replacing $c$ with $\mathscr{U}$ and $\mu$ with $T$) temperature driven martensitic transformations for solids that exhibit instabilities with respect to both internal energy density and strain. Past treatments of this phenomenon also relied on a Barsch and Krumhansl approach, solving the mechanical problem together with Fourier's law of heat conduction  \cite{Bouville,GarthWells}. However, these studies were restricted to two dimensions \cite{Bouville,GarthWells,AlphonseFinel} with some neglecting geometric non-linearity by relying on infinitesimal strains \cite{Bouville,GarthWells}. At a more fundamental level, they did not consider the possibility of spinodal instabilities with respect to the redistribution of internal energy density.

\section{Methods} The governing fourth-order PDEs of mass transport and gradient elasticity are solved numerically in weak form, wherein second-order spatial gradients appear on the trial solutions and their variations. Numerical solutions require basis functions that are continuous up to their first spatial gradients, at least. Our numerical framework (see Rudraraju et al. \cite{Rudrarajuetal2014}) uses isogeometric analysis \cite{CottrellHughesBazilevs2009} and the PetIGA code framework \cite{VictorCalo} with spline basis functions, which can be constructed for arbitrary degree of continuity (Supporting Information). This framework has proven pivotal to the current work. The code used for the numerical examples in this paper can be downloaded at the University of Michigan Computational Physics Group open source codes webpage: \url{http://www.umich.edu/~compphys/codes.html}
 
\section{Funding}
The mathematical formulation for this work was carried out under an NSF CDI Type I grant: CHE1027729 ``Meta-Codes for Computational Kinetics'' and NSF DMR 1105672. The numerical formulation and computations have been carried out as part of research supported by the U.S. Department of Energy, Office of Basic Energy Sciences, Division of Materials Sciences and Engineering under Award \#DE-SC0008637 that funds the PRedictive Integrated Structural Materials Science (PRISMS) Center at University of Michigan.

\end{document}


\title{Supplementary Material for Mechano-chemical spinodal decomposition: A phenomenological theory of phase transformation in multi-component, crystalline solids}
\author{Shiva Rudraraju\thanks{Mechanical Engineering, University of Michigan} , Anton Van der Ven\thanks{Materials, University of California Santa Barbara}, 
Krishna Garikipati\thanks{Mechanical Engineering, Mathematics, University of Michigan, \href{mailto:krishna@umich.edu}{\nolinkurl{krishna@umich.edu}}}}
\maketitle

\beginsupplement

\section{The existence of a free energy density function that is non-convex with respect to composition}

Consider the experimentally derived equilibrium phase diagram for transition metal hydrides \cite{Zuzeketal1990} shown schematically in Figure 2c of the main text. The single cubic phase, stable for all composition and strains at high temperature, only remains stable at high composition when quenched to low temperature. At low composition, the tetragonal phase is stable at low temperature. Also recall that first principle calculations have shown that the tetragonal phase is unstable with respect to strain and decomposes into three energetically equivalent variants \cite{ThomasAVDV}. The homogeneous free energy density therefore is symmetric with respect to zero strain. The two-phase coexistence region implies that the free energy density surface, $\mathscr{F}$ in strain-composition, $(\be,c)$ space admits a common tangent hyperplane. For clarity, let us consider the zero strain hyperplane in this space. Its projection appears in Figure S1. For a point in the solid with composition $c^\ast$, the common tangent construction leads to compositions $c_\alpha$ and $c_\beta$, respectively of the tetragonal and cubic phases, with 
\begin{equation}
\frac{\partial\mathscr{F}}{\partial c}\Bigg\vert_{c_\alpha} = \frac{\partial\mathscr{F}}{\partial c}\Bigg\vert_{c_\beta}.
\label{eq:commontang}
\end{equation}
Furthermore, the stability of the cubic and tetragonal phases with respect to composition imply that
\begin{equation}
\frac{\partial^2\mathscr{F}}{\partial c^2}\Bigg\vert_{c_\alpha} > 0,\quad \frac{\partial^2\mathscr{F}}{\partial c^2}\Bigg\vert_{c_\beta} >0.
\label{eq:poscurv}
\end{equation}
Recognizing that $c_\beta > c_\alpha$ in Figure 2c of the main text, it follows from Equation (\ref{eq:poscurv}) that there exists $c_\alpha^+ > c_\alpha$ and $c_\beta^- < c_\beta$ such that 
\begin{equation}
\frac{\partial\mathscr{F}}{\partial c}\Bigg\vert_{c_\alpha^+} > \frac{\partial\mathscr{F}}{\partial c}\Bigg\vert_{c_\alpha}, \qquad\mathrm{and}\; \frac{\partial\mathscr{F}}{\partial c}\Bigg\vert_{c_\beta^-} <  \frac{\partial\mathscr{F}}{\partial c}\Bigg\vert_{c_\beta}.
\label{eq:convex}
\end{equation}
But then, from Equations (\ref{eq:commontang}) and (\ref{eq:convex}), it follows that, for a smooth $\mathscr{F}$ that is at least twice differentiable in the interval $(c_\alpha,c_\beta)$, there must exist at least one value $c_\gamma$ with $c_\alpha^+ \le c_\gamma \le c_\beta^-$ satisfying
\begin{equation}
\frac{\partial^2\mathscr{F}}{\partial c^2}\Bigg\vert_{c_\gamma} < 0.
\label{eq:nonconvex}
\end{equation}
Therefore there is at least one intermediate value of composition in the two-phase coexistence region at which the free energy density is non-convex with respect to composition. This result is simply a consequence of the mean value theorem of calculus.

\begin{figure}[h]
  \centering
  \includegraphics[width=0.5\textwidth]{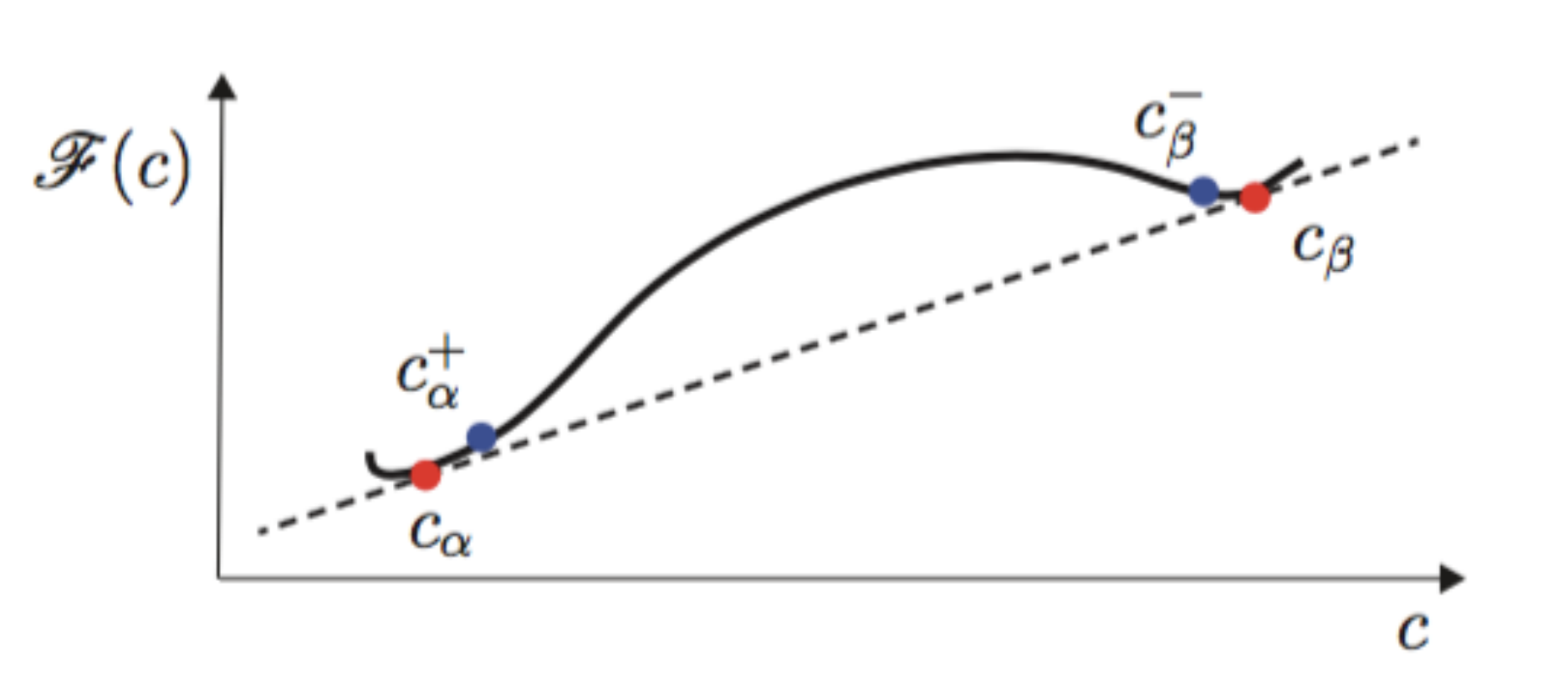}
\caption{\textbf{Figure S1}: Projection of the free energy on to the zero strain hyperplane.}
\label{fig:commtangent}
\end{figure}

\section{Finite strain kinematics}
In the interest of clarity and consistency with the main text, the Euclidean basis of our Cartesian coordinate system is denoted by $\bX_i$, $i = 1,\dots 3$, $\bX_i\cdot \bX_j = \delta_{ij}$. This constitutes a mild abuse of notation, because reference positions have also been denoted by $X_1, X_2, X_3$. The high symmetry, cubic reference configuration of the crystal is denoted by $\Omega$, on which material points are labelled by their position vectors $\bX = X_1\bX_1 + X_2\bX_2 + X_3\bX_3$. The deformation map  from $\Omega$ to the current configuration of the crystal is $\Bvarphi(\bX,t) =  \bX + \bu$, where the displacement field $\bu$ was introduced in the main text. The deformation gradient tensor is $\bF = \partial\Bvarphi/\partial\bX = \bone + \partial\bu/\partial\bX$. Using lower case indices to denote the components of vectors and tensors on the current configuration, and upper case indices for those on the reference configuration, we have the coordinate expression $F_{iJ} = \partial\varphi_i/\partial X_J=\delta_{iJ}+\partial u_{i}/\partial X_J$. 

The Green-Lagrange strain tensor is  $\bE = \frac{1}{2}(\bF^\mathrm{T}\bF - \bone)$, where $\bone$ is the second-order isotropic tensor. We repeat this relation in coordinate notation from the main text: $E_{IJ} = \frac{1}{2}(F_{iI}F_{iJ} - \delta_{IJ})$. The polar decomposition theorem states that the deformation gradient can be decomposed multiplicatively as $\bF = \bR\bU$, where $\bR$ is an (orthogonal) rotation tensor belonging to the space $\mathrm{SO}(3)$. It therefore satisfies $\bR^\mathrm{T}\bR = \bone$. The other tensor in this decomposition, $\bU$, is symmetric and positive-definite. From the polar decomposition it follows that $\bE$ as defined above is invariant to rotation of the current configuration. The requirement of material frame invariance dictates that the dependence of the free energy density on elasticity must be as a function of $\bE$. Such a free energy density function is invariant to rotations on the current configuration.

\section{The chemical potential: Variational derivation} 
Since the crystal is, in general, far from chemical equilibrium we follow the standard treatment of non-equilibrium thermodynamics to model the chemistry. In order to obtain a relation for the chemical potential we consider variations on the composition field of the form $c_\varepsilon := c + \varepsilon\,\xi$, where $\xi$ is a function such that, on the Dirichlet boundary, $\Gamma_c \subset \Gamma$, we have $c =\bar{c}$ and  $\xi = 0$.
\begin{align}
\frac{\delta}{\delta c}\Pi[c , \bu] = &\frac{d \Pi[c  + \varepsilon\xi, \bu]}{d\varepsilon}
\bigg|_{\varepsilon=0} \nonumber \\
\end{align}

\noindent Carrying out the differentiation with respect to $\varepsilon$ and integration by parts yields
\begin{align}
\frac{\delta\Pi}{\delta c}  = \int\limits_{\Omega}  \xi  &\Bigg(\frac{\partial \mathscr{F}}{\partial c} - \nabla \cdot (\Bkappa^\mathrm{s} ~\nabla  c )+ \nabla c\cdot\frac{\partial \Bkappa}{\partial c}\nabla c \nonumber\\
&+  \sum\limits_{\alpha,\beta}\frac{1}{2}\nabla e_\alpha\cdot\frac{\partial\Bgamma^{\alpha\beta}}{\partial c}\nabla e_\beta\nonumber\\
&+ \sum\limits_{\alpha}\left(\nabla c\cdot\frac{\partial\Btheta^\alpha}{\partial c}\nabla e_\alpha -\nabla\cdot\left(\Btheta^\alpha\nabla e_\alpha\right)\right)\Bigg)  ~\mathrm{d}V \nonumber\\
&+ \int\limits_{\Gamma}\xi \,\left(\Bkappa^\mathrm{s}\nabla c+\sum\limits_\alpha\Btheta^\alpha\nabla e_\alpha \right)\cdot\bn\, \mathrm{d}S,
  \label{eqn:formulation:variationChemistry}
\end{align}
  
\noindent where $\bn$ is the outward normal vector to $\Gamma$. At chemical equilibrium we have $\delta\Pi/\delta c = 0$, yielding two conditions:
\begin{align}
\int\limits_{\Omega}  \xi  &\Bigg[\frac{\partial \mathscr{F}}{\partial c} - \nabla \cdot (\Bkappa^\mathrm{s} ~\nabla  c )+ \nabla c\cdot\frac{\partial \Bkappa}{\partial c}\nabla c\nonumber\\
&+ \sum\limits_{\alpha,\beta}\frac{1}{2}\nabla e_\alpha\cdot\frac{\partial\Bgamma^{\alpha\beta}}{\partial c}\nabla e_\beta\nonumber\\
&+ \sum\limits_{\alpha}\left(\nabla c\cdot\left(\frac{\partial\Btheta^\alpha}{\partial c}\nabla e_\alpha\right) -\nabla\cdot\left(\Btheta^\alpha\nabla e_\alpha\right)\right)\Bigg]  ~\mathrm{d}V = 0\label{chemeq1}\\
\int\limits_\Gamma \xi&\left[\Bkappa^\mathrm{s}\nabla c+\sum_\alpha\Btheta^\alpha\nabla e_\alpha \right]\cdot\bn = 0
\label{chemeq2}
\end{align}
Standard variational arguments lead to the conclusion that, for chemical equilibrium, the quantities within the brackets must vanish on the corresponding domains $\Omega$ and $\Gamma$. The expression contained within brackets in the integrand of Equation [\ref{chemeq1}] is the chemical potential $\mu$. In the non-equilibrium setting that is of interest here, we continue to be guided by the above treatment and use the same definition for the chemical potential \cite{degrootmazur1984}:
\begin{align}
\mu = &\frac{\partial \mathscr{F}}{\partial c} -\nabla\cdot\left(\Bkappa^\mathrm{s}\nabla c\right) + \nabla c\cdot\frac{\partial \Bkappa}{\partial c}\nabla c\nonumber\\
& + \sum\limits_{\alpha,\beta}\frac{1}{2}\nabla e_\alpha\cdot\frac{\partial\Bgamma^{\alpha\beta}}{\partial c}\nabla e_\beta\nonumber\\
&  + \sum\limits_{\alpha}\left(\nabla c\cdot\frac{\partial\Btheta^\alpha}{\partial c}\nabla e_\alpha - \nabla \cdot\left(\Btheta^\alpha\nabla e_\alpha\right)\right).
\label{chempot}
\end{align}

\noindent Variants of Equation [\ref{chemeq2}] appear in a proper variational derivation of chemical equilibrium---a detail that is sometimes overlooked in phase field treatments. We will impose chemical equilibrium of the boundary $\Gamma$, by requiring the term in brackets to vanish.

With the standard, phenomenological relation $\bJ = -\bL(c,\be)\nabla\mu$ for the flux, the strong form of the governing PDE of mass balance follows:
\begin{align}
\frac{\partial c}{\partial t} + \nabla\cdot\bJ &= 0 &\mathrm{in}\; \Omega &\nonumber\\
-\bJ\cdot\bn &= \overline{J} &\mathrm{on}\; \Gamma &\nonumber\\
(\Bkappa^\mathrm{s}\nabla c+\sum_\alpha\Btheta^\alpha\nabla e_\alpha )\cdot\bn &= 0 &\mathrm{on}\; \Gamma &\nonumber\\
c(\bX) &= c_0(\bX) &\mathrm{at}\;t=0  &
\label{strongformdiff}
\end{align}
\noindent Note that we have assumed that there is no Dirichlet boundary; i.e. $\Gamma_c = \emptyset$, in stating the final strong form of this PDE. The conjugate boundary condition on the mass flux holds on the entire boundary $\Gamma$, in Equation [\ref{strongformdiff}$_2$]. Additionally, we have the boundary condition arising from requiring chemical equilibrium on $\Gamma$, in Equation [\ref{strongformdiff}$_3]$. Using standard variational arguments, the weak form of mass balance can be obtained by starting from [\ref{strongformdiff}]. Finally, we have the initial condition on composition, Equation [\ref{strongformdiff}$_4$]. The substitution of Equation [\ref{chempot}] in $\bJ = -\bL\nabla\mu$, followed by this phenomenological flux relation's substitution in (\ref{strongformdiff}$_1$) shows that this is a fourth order PDE in space. 

\subsection{On mass transport posed in the reference configuration}
We note that the mathematical formulation of diffusion is commonly stated in the current (deformed) configuration, which is the true state of the solid. However, for crystalline solids (with negligible extended defects) it is more convenient to define Onsager transport coefficients and evaluate composition gradients on the reference (undeformed) configuration, here denoted by $\Omega$. Diffusion in the crystalline solids results from atomic hops between well-defined crystal sites, which can always be mapped onto a reference crystal structure. Increasingly, mobility coefficients in multi-component solids are calculated from first principles through the evaluation of Kubo-Green expressions in kinetic Monte Carlo simulations on the reference crystal in configuration $\Omega$. Since the mobility tensor is reported on $\Omega$, it is most convenient to adopt the Lagrangian description for diffusion over this configuration in single crystalline solids.

Our numerical implementation is based on the weak form, whose derivation we begin by first reintroducing $\xi$, now in the guise of a weighting function lying in the function space $\mathscr{Q} = \{\xi\,\in\, H^1(\Omega)\}$. We seek functions $c\, \in\, \mathscr{P} = \{c\,\in\, H^1(\Omega)\}$ such that
\begin{align}
  \int\limits_{\Omega}  \xi \frac{\partial  c }{\partial t} \mathrm{d}V &+ \int\limits_{\Omega} \nabla\xi \cdot\bL \nabla \left( \frac{\partial \mathscr{F}}{\partial c}  + \frac{\partial \mathscr{G}}{\partial c} \right)\mathrm{d}V\nonumber\\
  &+\int\limits_\Omega\nabla\cdot(\bL\nabla\xi) \nabla\cdot(\Bkappa\nabla c) \mathrm{d}V\nonumber\\
  & -\int\limits_{\Gamma}\,\xi\overline{J} \mathrm{d}S- \int\limits_{\Gamma}  (\bL\nabla\xi)\cdot\bn \,\nabla\cdot(\Bkappa\nabla c) \mathrm{d}S = 0
\label{weakmasstransport}
\end{align}
Equation [\ref{weakmasstransport}] is obtained by multiplying $\xi$ into [\ref{strongformdiff}$_1$], integrating by parts twice, and using the Neumann boundary condition [\ref{strongformdiff}$_2$]. Note that the higher-order Dirichlet boundary condition (\ref{strongformdiff}$_3$) has not been built into the function spaces as is commonly done in weak forms. Instead, it is imposed weakly using the method of Nitsche \cite{Nitsche1971,Arnoldetal2001,BazilevsHughes2007} and extending the weak form to
\begin{align}
  \int\limits_{\Omega}  \xi \frac{\partial  c }{\partial t} \mathrm{d}V &+ \int\limits_{\Omega} \nabla\xi \cdot\bL \nabla \left( \frac{\partial \mathscr{F}}{\partial c}  + \frac{\partial \mathscr{G}}{\partial c} \right)\mathrm{d}V\nonumber\\
  &+\int\limits_\Omega\nabla\cdot(\bL\nabla\xi) \nabla\cdot(\Bkappa\nabla c)  \mathrm{d}V\nonumber\\
  & -\int\limits_{\Gamma}\,\xi\overline{J} \mathrm{d}S- \int\limits_{\Gamma}  (\bL\nabla\xi)\cdot\bn \,\nabla\cdot(\Bkappa\nabla c) \mathrm{d}S\nonumber\\
  & - \int\limits_{\Gamma}   \nabla\cdot(\Bkappa\nabla \xi) \,(\bL\nabla c)\cdot\bn\mathrm{d}S\nonumber\\
  &  + \int\limits_\Omega \tau \nabla\xi\cdot\bn \nabla c\cdot\bn\mathrm{d}S  = 0
\label{weakmasstransportsymm}
\end{align}

Equation [\ref{weakmasstransportsymm}] is based on the coupling tensor $\Btheta^\alpha = \bzero$ for all $\alpha$, and for $\Bkappa, \Bgamma^{\alpha\beta}$ independent of $c$, corresponding to the numerical examples presented in the main text. This weak form maintains consistency. Note that the second-to-last term that has been added over the form in Equation [\ref{weakmasstransport}] is symmetric with the term preceding it. This formulation shows better numerical performance than the also-consistent formulation in which this additional term is preceded by a negative sign. Finally, $\tau$ is a stabilization parameter that is inversely proportional to the element size.  

\section{Mechanical equilibrium}
As explained in the main text, since elastic wave propagation is orders of magnitude faster than the kinetics of mass transport, it can be assumed that mechanical equilibrium is attained instantaneously for the prevailing composition field. To obtain the relevant equilibrium equations, we introduce variations on the displacement field: $\bu_\varepsilon = \bu + \varepsilon\bw$, such that on the Dirichlet boundary $\Gamma_{u_i}$, the fields satisfy $u_i = \bar{u}_i$ and $w_i = 0$. We construct the first variation of $\Pi[c, \bu]$ with respect to $\bu$.

\begin{align}
\frac{\delta}{\delta\bu}\Pi[ c , \bu] = \frac{d}{d\varepsilon}\Bigg(\int\limits_{\Omega}\left(\mathscr{F}(c,\be_{\varepsilon}) + \mathscr{G}(c,\be_\varepsilon,\nabla c,\nabla\be_\varepsilon)\right)    ~\mathrm{d}V&\nonumber\\ 
- \sum\limits_{i=1}^{3}\int\limits_{\Gamma_{T_i}}u_{i_\varepsilon}T_i \mathrm{d}S\Bigg) \Bigg\vert_{\varepsilon=0}&
\label{varderivativemech}
\end{align}

\noindent where $\be_{\varepsilon}$  are obtained by first replacing $\bu$ by $\bu_\varepsilon$ in the kinematic relations for $\bE$ discussed above, which are then substituted in the relations for the reparameterized strains $\be$. We recall the constitutive relations for the first Piola-Kirchhoff stress and the higher-order stress, respectively:
\begin{align}
P_{iJ}&= \sum\limits_{\alpha}\frac{\partial (\mathscr{F}+\mathscr{G})}{\partial e_\alpha}  \frac{\partial e_\alpha}{\partial F_{iJ}} +  \sum\limits_{\alpha}\frac{\partial \mathscr{G}}{\partial e_{\alpha,I}}\frac{\partial e_{\alpha,I}}{\partial F_{iJ}} \label{stressP}\\
B_{iJK} &= \sum\limits_{\alpha}\frac{\partial \mathscr{G}}{\partial e_{\alpha,I}}  \frac{\partial e_{\alpha,I}}{\partial F_{iJ,K}}\label{stressB}
\end{align}

\noindent Using these constitutive relations, Equation  [\ref{varderivativemech}] can be manipulated to yield the Euler-Lagrange equation, which is also the weak form of mechanical equilibrium for a strain gradient elastic material, on $\Omega$ in terms of $\bP$ and $\bB$. We retain coordinate notation in the interest of transparency:
\begin{equation}
  \int\limits_{\Omega} \left( P_{iJ} w_{i,J} +  B_{iJK} w_{i,JK} \right) ~\mathrm{d}V -\sum\limits_{i=1}^{3} \int\limits_{\Gamma_{T_i}} w_i T_i \, ~\mathrm{d}S  = 0
  \label{weakformmech}
\end{equation}

\noindent We introduce the normal and surface gradient operators, $\nabla^n$ and $\nabla^s$, defined by
\begin{equation}
\nabla^n \psi = \nabla\psi\cdot\bn, \;\mathrm{and}\;\nabla^s\psi = \nabla\psi - (\nabla^n\psi)\bn.
\label{surfacenormalgradient}
\end{equation}

\noindent Starting from Equation [\ref{weakformmech}] and applying integration by parts several times we have the strong form of mechanical equilibrium for a strain gradient elastic material. See Rudraraju et al. \cite{Rudrarajuetal2014} for detailed steps. The results of this variational treatment are analogous to those of Toupin \cite{Toupin1962}, except that it is carried out over the reference configuration of the crystal, rather than its current configuration.
\begin{align}
P_{iJ,J} - B_{iJK,JK} &= 0 \;\;\;\mathrm{in} \;\Omega\nonumber\\
u_{i}  &= \bar{u}_i  \; \mathrm{on} \;\Gamma_{u_i}\nonumber\\
P_{iJ}n_J - \nabla^n B_{iJK} n_K n_J - 2\nabla^s_J(B_{iJK})n_K & \nonumber\\
- B_{iJK}\nabla^s_J \nabla^s_K + (b_{LL}n_J n_K-b_{JK})B_{iJK} & = T_{i}\; \mathrm{on} \;\Gamma_{T_i}\nonumber\\
B_{iJK}n_J n_K &= 0 \;\;\mathrm{on}  \;\Gamma\nonumber\\
\llbracket n^{\Gamma}_{J} n_{K} B_{iJK} \rrbracket &= 0 \;\;\mathrm{on} \;\Upsilon_{L_i}\label{strongformgradelasticity}
\end{align}
Here, $\Gamma= \Gamma_{u_i} \cup  \Gamma_{T_i}$ for $i = 1,2,3$, is the decomposition of the boundary surface into a subset with Dirichlet boundary condition on displacement component $u_i$ and Neumann boundary condition on traction component $T_i$, respectively. Finally, $\Upsilon$ is a smooth boundary edge on which a jump in higher-order stress traction may arise. The fourth-order nature of the governing partial differential equation emerges on evaluating [\ref{stressP}--\ref{stressB}] and substituting in [\ref{strongformgradelasticity}$_1$]. 

The Dirichlet boundary condition in [\ref{strongformgradelasticity}$_2$] has the same form as for conventional elasticity. However, its dual Neumann boundary condition, [\ref{strongformgradelasticity}$_3$] is notably more complex than its conventional counterpart, which would have only the first term on the left hand-side. Equation [\ref{strongformgradelasticity}$_4$] is the higher-order Neumann boundary condition on the higher-order stress traction. The physical interpretation of this boundary condition in its homogeneous form is that the boundary has no mechanism to impose a moment on bonds at the atomic scale. Equation [\ref{strongformgradelasticity}$_5$] specifies that there is no discontinuity of higher-order stress traction across edges. The weak form of mechanical equilibrium has already been encountered in [\ref{weakformmech}]. For completeness we specify functional spaces: Find $\bu$ lying in $\mathscr{S} = \{\bu\;\in\;H^2(\Omega_0)\vert\;u_i = \bar{u}_i\;\mathrm{on}\;\Gamma_{u_i}\}$ such that given functions $T_i$ on $\Gamma_{T_i}$ for $ i = 1,2,3$ and tensors $\bP,\;\bB$ defined by Equations [\ref{stressP}--\ref{stressB}], Equation [\ref{weakformmech}] is satisfied for all $\bw$ lying in $\mathscr{V} = \{\bw\;\in\;H^2(\Omega)\vert\;w_i = 0\;\mathrm{on}\;\Gamma_{u_i}\}$. 

\section{Isogeometric Analysis for high-order PDEs}
\label{sec:iga}
Isogeomeric Analysis (IGA) is a mesh-based numerical method with NURBS (Non-Uniform Rational B-Splines) basis functions. The NURBS basis functions have many desirable properties. They are partitions of unity with compact support, satisfy affine covariance and provide certain advantages over Lagrange polynomial basis functions, which are the mainstay of Galerkin finite element methods. These advantages include the imposition of higher-order, $C^n$-continuity, positivity, convex hull properties, and being variation diminishing. For a detailed discussion of the NURBS basis and IGA, interested readers are referred to  Cottrell et al. \cite{CottrellHughesBazilevs2009}. However, we briefly present the construction of the basis functions. \\

\begin{figure}[hbt]
  \centering
  \includegraphics[width=0.5\textwidth]{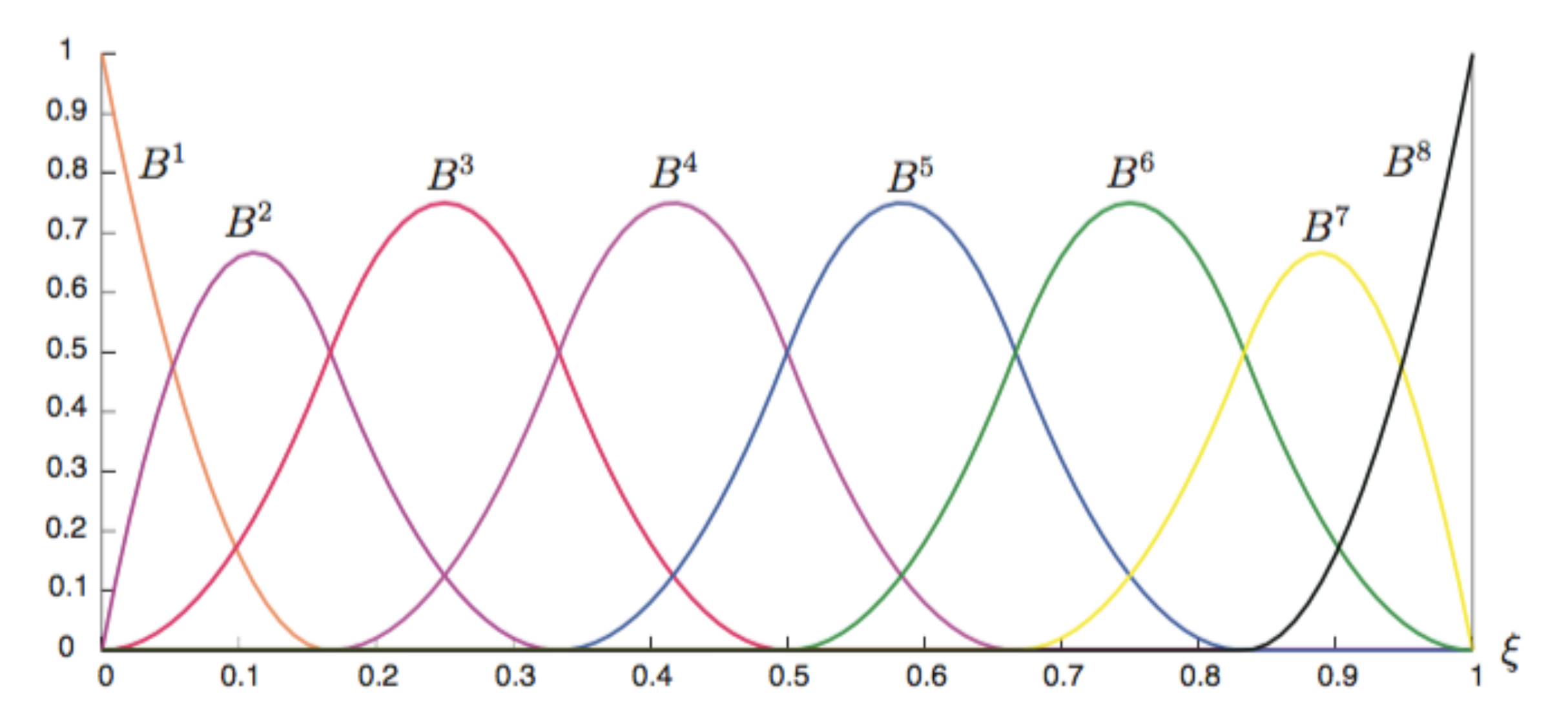}
\caption{\textbf{Figure S2}: Quadratic (p=2) B-spline basis constructed from the knot vector $\chi=$ [0, 0 , 0 , 1/6, 1/3, 1/2, 2/3, 5/6, 1, 1, 1].}
\label{fig:bsplines}
\end{figure}

The building blocks of the NURBS are univariate B-spline functions that are defined as follows: Consider positive integers $p$ and $n$, and a non-decreasing sequence of values $\chi=[\xi_1, \xi_2,...., \xi_{n+p+1}]$, where p is the polynomial order, n is the number of basis functions, the $\xi_i$ are coordinates in the parametric space referred to as knots (equivalent to nodes in FEM) and $\chi$ is the knot vector. The B-spline basis functions $B_{i,p}(\xi)$ are defined starting with the zeroth order basis functions
\begin{align}
B_{i,0}(\xi) &= \left\{\begin{array}{ll}
1 &\mathrm{if}\;\xi_i \le \xi < \xi_{i+1},\\
0 &\mathrm{otherwise}
\end{array}\right.
\end{align}
and using the Cox-de Boor recursive formula for $p \geq 1$ \cite{Piegl1997}
\begin{align}
 B_{i,p} (\xi) &=
  \frac{\xi-\xi_i}{\xi_{i+p}-\xi_i} B_{i,p-1} (\xi) + \frac{\xi_{i+p+1}-\xi}{\xi_{i+p+1}-\xi_{i+1}} B_{i+1,p-1} (\xi)
\end{align}
The knot vector divides the parametric space into intervals referred to as knot spans (equivalent to elements in FEM). A B-spline basis function is $C^{\infty}$-continuous inside knot spans and $C^{p-1}$-continuous at the knots. If an interior knot value repeats, it is referred to as a multiple knot. At a knot of multiplicity $k$, the continuity is $C^{p-k}$. The boundary value problems in the main text consider only simple geometries, for which B-spline basis functions have sufficed. However, we outline the extension to NURBS basis functions for the sake of completeness, noting that the numerical formulation as presented is valid for any single-patch NURBS geometry. Using a quadratic B-spline basis (Figure \ref{fig:bsplines}), a $C^1$-continuous one dimensional NURBS basis can be constructed. 
\begin{align}
N^{i} (\xi) =
  \frac{B_{i,2} (\xi) \textit{w}_{i}}{\sum_{i=1}^{n_b} B_{i,2} (\xi) \textit{w}_{i}}
  \label{eq:onednurbs}
\end{align}
where $w_i$ are the weights associated with each of the B-spline functions. In higher-dimensions, NURBS basis functions are constructed as a tensor product of the one dimensional basis functions:
\begin{align}
 N^{ij} (\xi,\eta) &=
  \frac{B_{i,2} (\xi) B_{j,2} (\eta) \textit{w}_{ij}}{\sum\limits_{i=1}^{n_{b1}} \sum\limits_{j=1}^{n_{b2}} B_{i,2}(\xi) B_{j,2} (\eta) \textit{w}_{ij}} &\mathrm{(2D)} \nonumber \\
 N^{ijk} (\xi,\eta, \zeta) &=
  \frac{B_{i,2} (\xi) B_{j,2} (\eta) B_{k,2} (\zeta) \textit{w}_{ijk}}{\sum\limits_{i=1}^{n_{b1}} \sum\limits_{j=1}^{n_{b2}} \sum\limits_{k=1}^{n_{b3}} B_{i,2}(\xi) B_{j,2} (\eta) B_{k,2} (\zeta) \textit{w}_{ijk}} &\mathrm{(3D)}
\label{eq:higherordernurbs}
\end{align}

\section{Numerical examples}
\label{sec:numerical}
Recall that in two dimensions, the free energy density function depends only on $c, e_1, e_2$ and $e_6$. In this first communication we wish to focus only on broad phenomenology, for which reason we choose the simplest versions of the free energy density function. Apart from the  dependence on $c$, which is essential for the mechano-chemical spinodal, the coefficients are constant. We have introduced the coefficients $d_c$ and $d_e$, which control the depths of the free energy wells and $s_e$, which determines the strain minimum in terms of $e_2$. Figure 2 of the paper depicts this construction. We also reduce $\Bkappa$ and $\Bgamma^{\alpha\beta}$ to isotropic forms: $\Bkappa = \kappa\bone$, $\gamma^{\alpha\beta}_{IJ} = \lambda_{e} l_\mathrm{e}^2 \delta^{\alpha\beta}\delta_{IJ}$, and set the composition gradient-strain gradient coupling coefficients $\Btheta^\alpha = \bzero$. While the fullest possible complexity of the coupling is not revealed by these simplifications, the aim here is to present the essential physics that is universal to mechano-chemical spinodal decomposition, postponing details of the more complex couplings and tensorial forms to future communications, where they will be derived for specific materials systems. Accordingly, in two dimensions, we have:
\begin{align}
 \mathscr{F}\left( c,e_1,e_2,e_6\right) =&  16 d_c c^4 - 32 d_c c^3 + 16 d_c c^2 \nonumber\\
	& + \frac{2 d_e}{s_e^2}(e_1^2 + e_6^2 )\nonumber \\
	& +\frac{d_e}{s_e^4}e_2^4 + (2c-1) \frac{2 d_e}{s_e^2}e_2^2\label{2Dhomogfreeenergy}\\
	\mathscr{G}(\nabla c, \nabla e_2) =& \frac{1}{2}\nabla c\cdot \kappa\nabla c +\frac{1}{2}\nabla e_2\cdot \lambda_{e} l_\mathrm{e}^2 \nabla e_2 \label{2Dinhomogfreeenergy}
\end{align}	
where $d_c=2.0$, $d_e=0.1$, $s_e=0.1$, $\kappa=10^{-6}$ and $\lambda_{e}=10^{-7}$. The problem domain is a square of dimensions $0.01 \times 0.01$. \\ 
\vspace{0.5cm}\\
\noindent In keeping with the above-explained simplifications aimed at presenting the essential physics of mechano-chemical spinodal decomposition, the free energy density function used for the three-dimensional examples is:
\begin{align}
	\mathscr{F}\left( c,\be\right) = &16 d_c c^4 - 32 d_c c^3 + 16 d_c c^2 \nonumber\\
	&+ \frac{3 d_e}{2 s_e^2}(e_1^2 + e_4^2 + e_5^2  + e_6^2) \nonumber\\
	&+ \frac{3 d_e}{2 s_e^4} (e_2^2+e_3^2)^2 \nonumber\\
	&+ (1-c) \frac{d_e}{s_e^3} e_3 (e_3^2-3e_2^2) \nonumber\\
	&+ (2c-1) \frac{3 d_e}{2 s_e^2} (e_2^2+e_3^2) \label{3Dhomogfreeenergy}\\
	\mathscr{G}\left( c,\be, \nabla c, \nabla e\right) = &\frac{1}{2}\nabla c\cdot \kappa \nabla c +\frac{1}{2}\nabla e_2\cdot \lambda_{e} l_\mathrm{e}^2  \nabla e_2  \nonumber\\
	&+\frac{1}{2}\nabla e_3\cdot \lambda_{e} l_\mathrm{e}^2 \nabla e_3 \label{3Dinhomogfreeenergy}
\end{align}
where $d_c=2.0$, $d_e=100.0$, $s_e=0.1$, $\kappa=10^{-2}$ and $\lambda_{e}=1.0$. The problem domain is a unit cube. The strain gradient energy coefficient tensors have been chosen to be isotropic, $\gamma^{\alpha\beta}_{IJ} = \lambda_{e} l_\mathrm{e}^2 \delta^{\alpha\beta}\delta_{IJ}$, and the three-dimensional formulation of the main text has been simplified here by setting the composition gradient-strain gradient coupling coefficients $\theta^\alpha_{IJ} = 0$.

\section{List of videos}
\label{sec:movies}
\begin{figure}[hbt]
  \centering
  \includegraphics[width=0.5\textwidth]{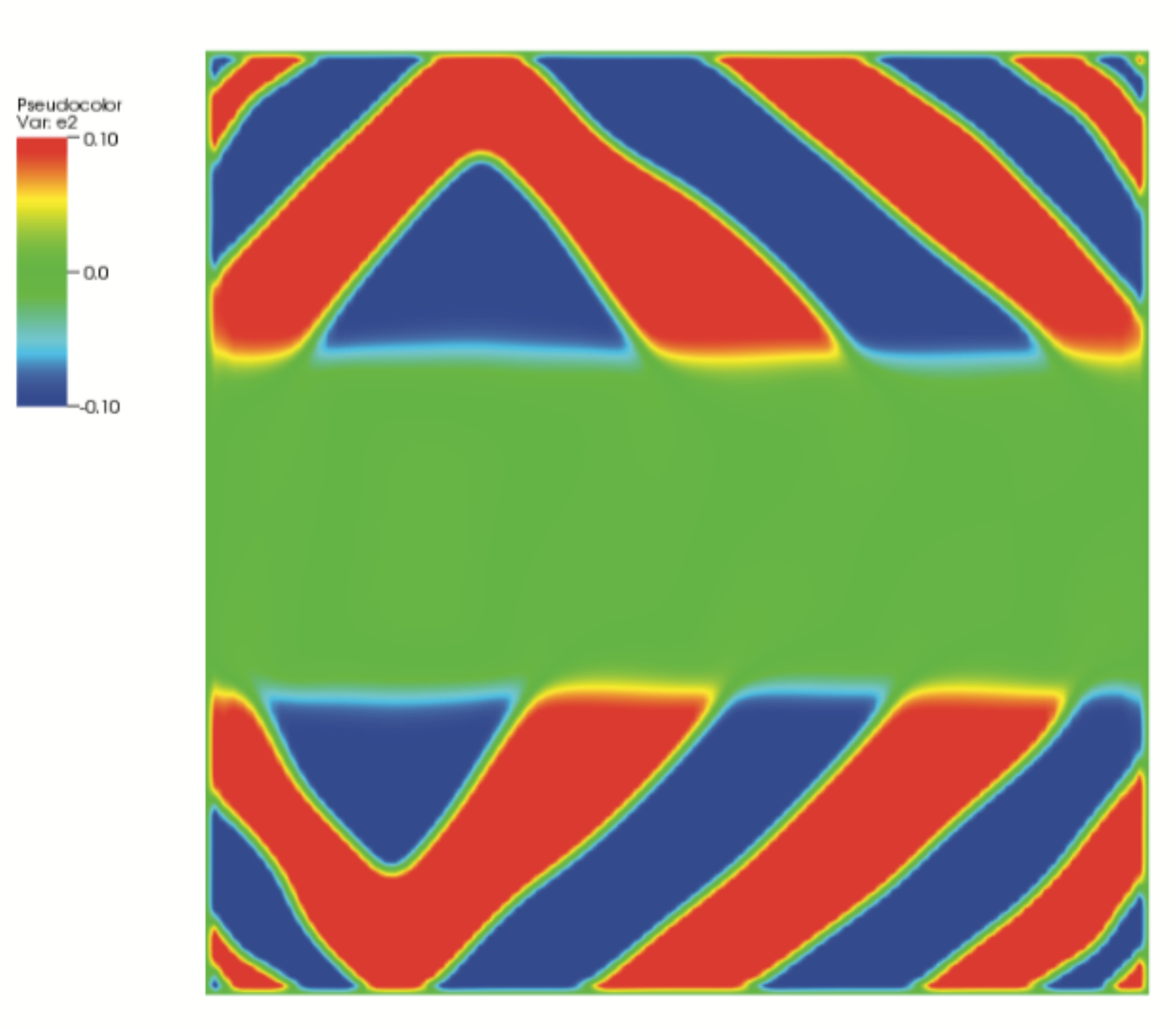}
\caption{\textbf{Movie S1}[MovieS1.mp4]: Evolution of microstructure of the two-dimensional solid with outward flux corresponding to the plot for $l_\mathrm{e}^2=0.1$ considered in Figure 5a of the main text.}
\end{figure}

\begin{figure}[hbt]
  \centering
  \includegraphics[width=0.5\textwidth]{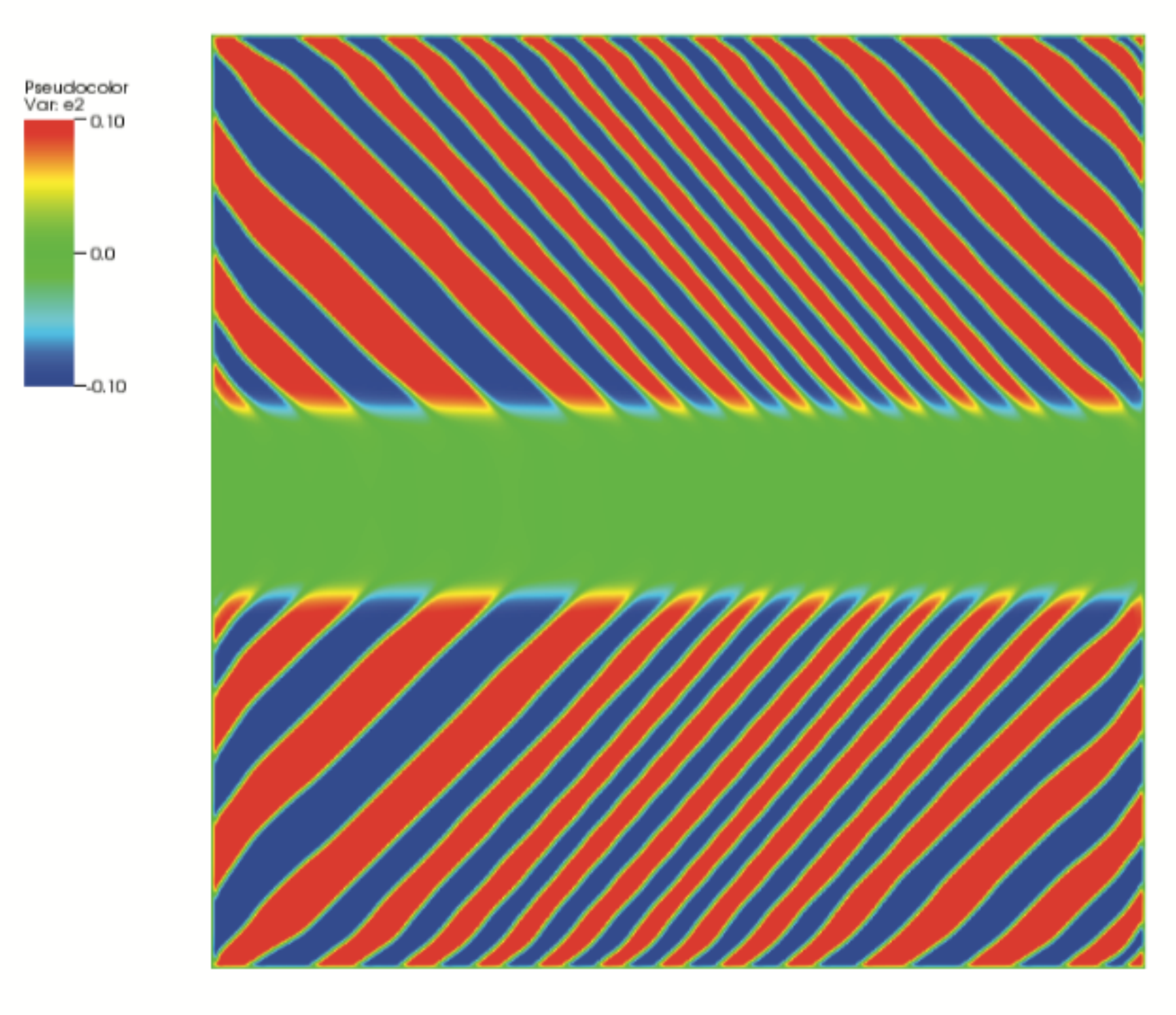}
\caption{\textbf{Movie S2}[MovieS2.mp4]: Evolution of microstructure of the two-dimensional solid with outward flux corresponding to the plot for $l_\mathrm{e}^2=0.01$ considered in Figure 5a of the main text.}
\end{figure}

\begin{figure}[hbt]
  \centering
  \includegraphics[width=0.5\textwidth]{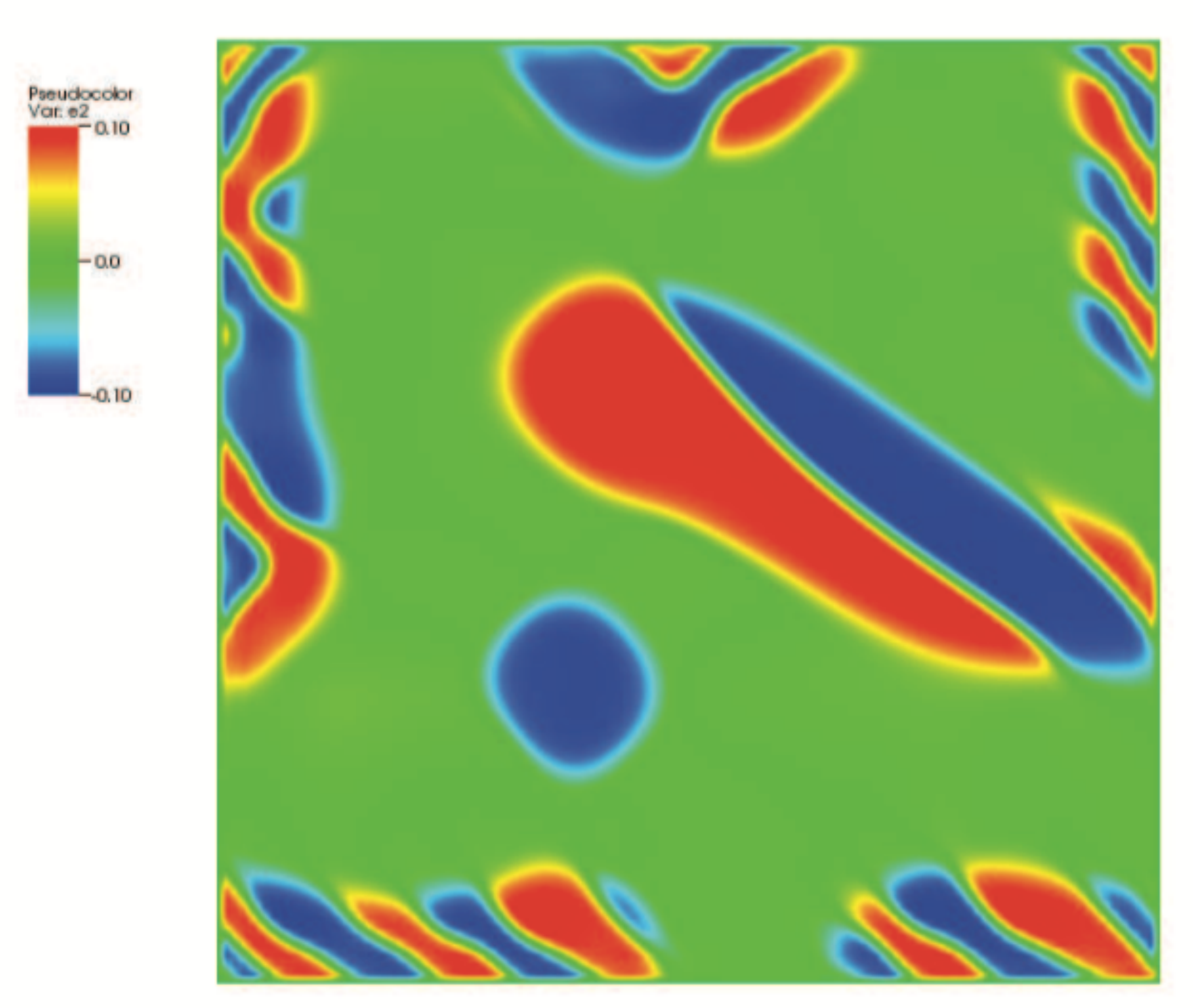}
\caption{\textbf{Movie S3}[MovieS3.mp4]: Evolution of microstructure of the two-dimensional solid due to quenching from an initial composition of $c = 0.46$. This corresponds to the plot for $l_\mathrm{e}^2=0.1$ considered in Figure 5b of the main text.}
\end{figure}
 
\begin{figure}[hbt]
  \centering
  \includegraphics[width=0.5\textwidth]{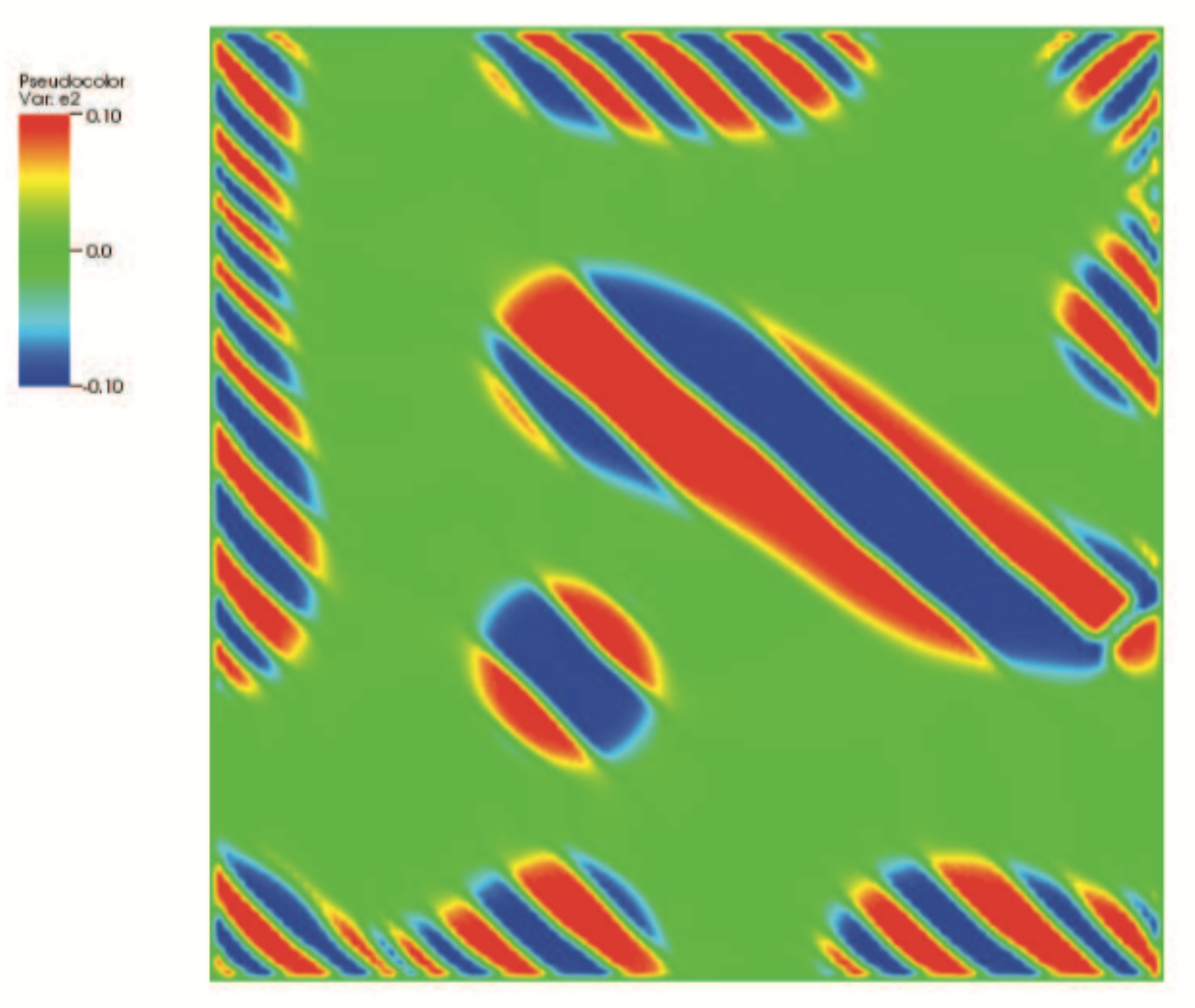}
\caption{\textbf{Movie S4}[MovieS4.mp4]: Evolution of microstructure of the two-dimensional solid due to quenching from an initial composition of $c = 0.46$. This corresponds to the plot for $l_\mathrm{e}^2=0.01$ considered in Figure 5b of the main text.}
\end{figure}

\begin{figure}[hbt]
  \centering
  \includegraphics[width=0.5\textwidth]{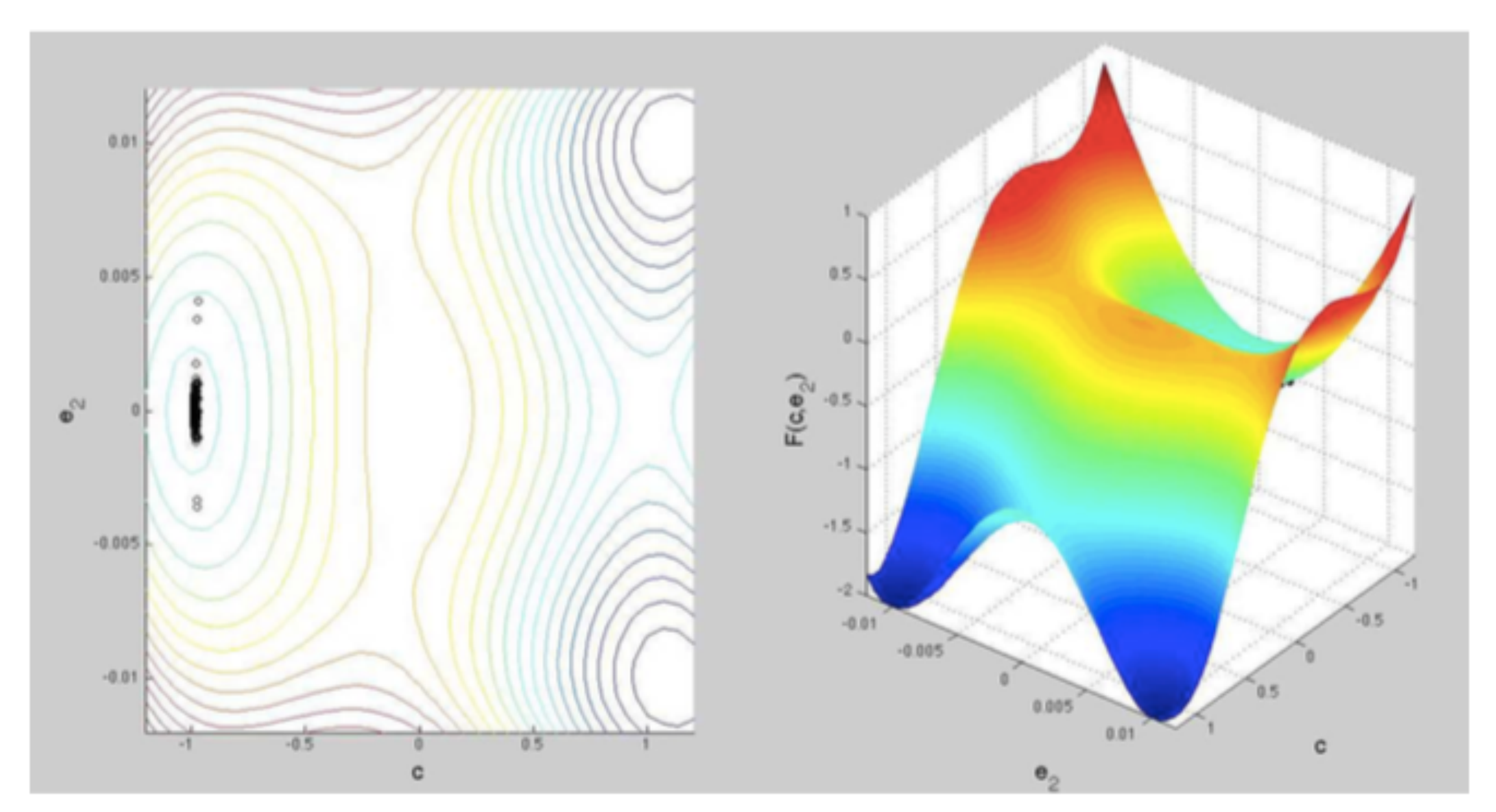}
\caption{\textbf{Movie S5}[MovieS5.mp4]: Evolution of material points on the homogeneous free energy density surface, $\mathscr{F}(c,e_1,e_2,e_6)$ for the two-dimensional formulation, restricted to the $\{c,e_2\}$ sub-space. The microstructure that forms is similar to that in Movie S1.}
\end{figure}

\begin{figure}[hbt]
  \centering
  \includegraphics[width=0.5\textwidth]{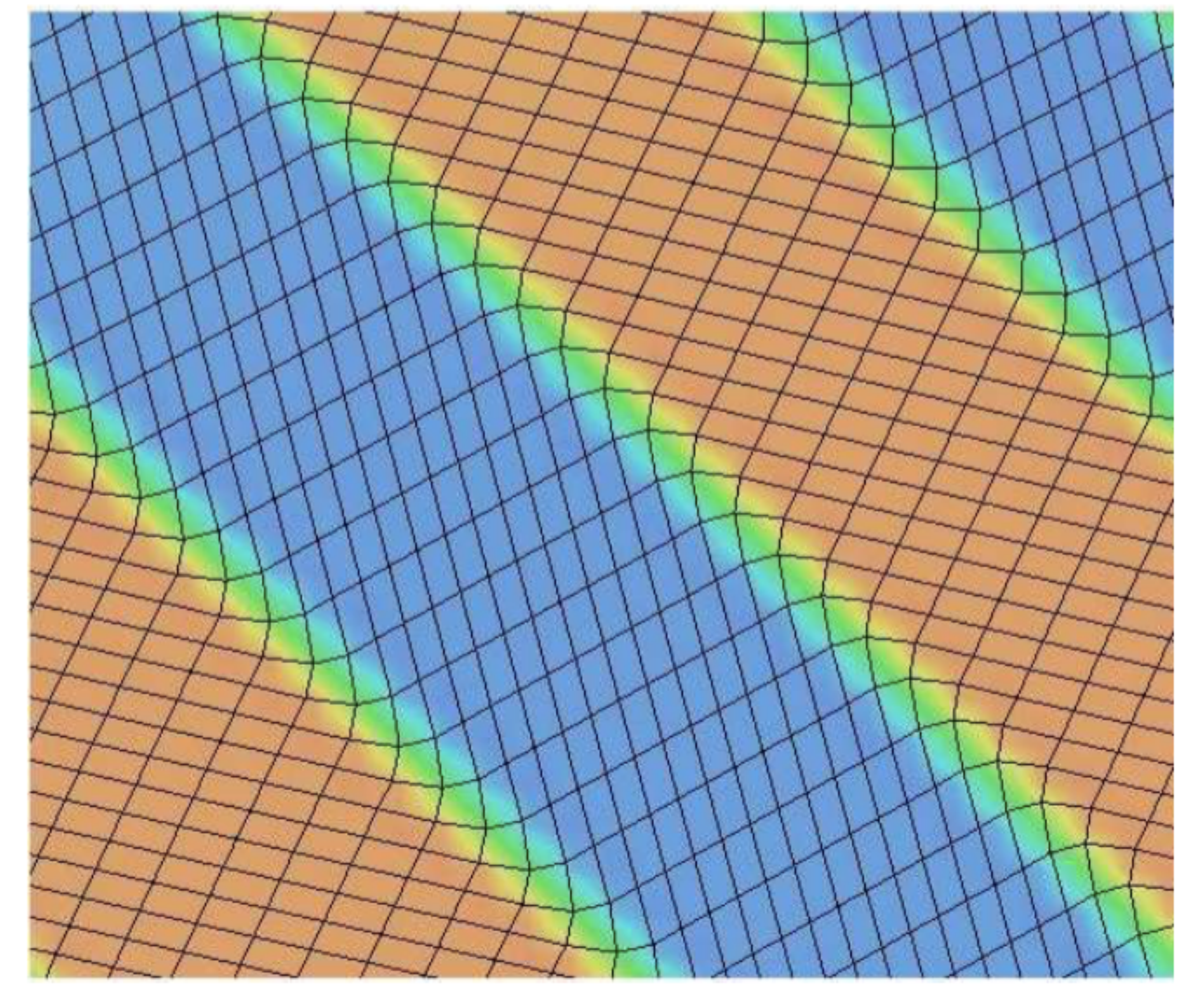}
\caption{\textbf{Movie S6}[MovieS6.mp4]: Zoomed-in view with distorted mesh lines showing the evolution of an initially square crystal lattice (green) into rectangular variants (red/blue) as the chemistry-driven cubic to rectangular transformation occurs, followed by the splitting of the variants into a twinned structure.}
\end{figure}

\begin{figure}[hbt]
  \centering
  \includegraphics[width=0.5\textwidth]{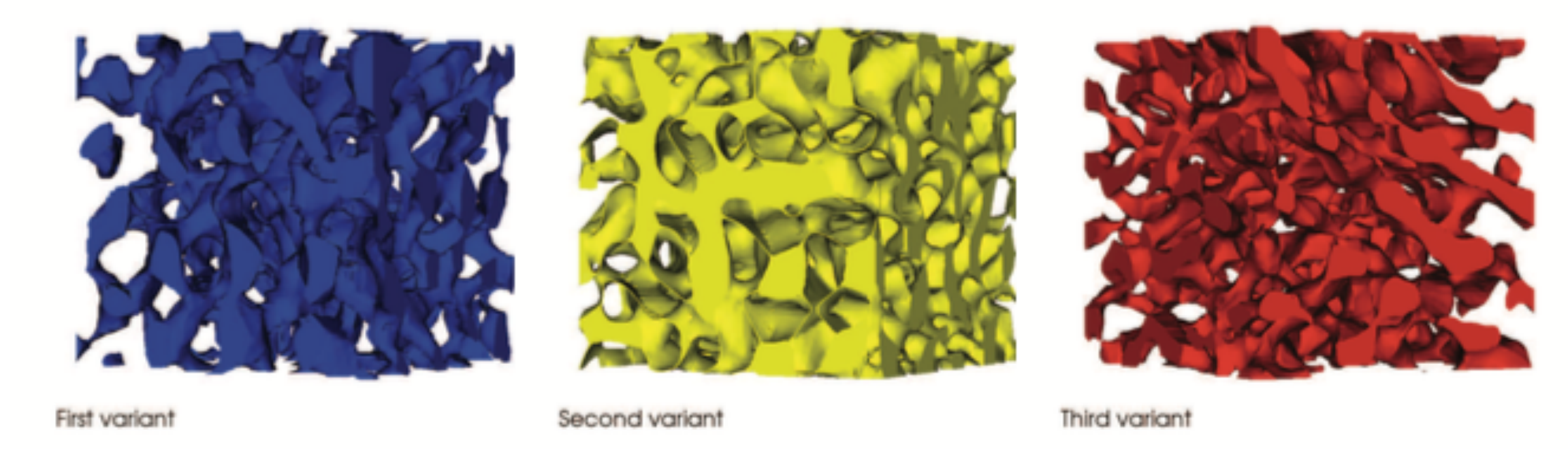}
\caption{\textbf{Movie S7} [MovieS7.mp4]: Isovolumes of the three interleaved tetragonal variants and the three-dimensional microstructure corresponding to the plot for $l_\mathrm{e}^2=1.0$ considered in Figure 6 of the main text. For visual clarity, only a sub-domain of the simulation volume is shown.}
\end{figure}

\begin{figure}[hbt]
  \centering
  \includegraphics[width=0.5\textwidth]{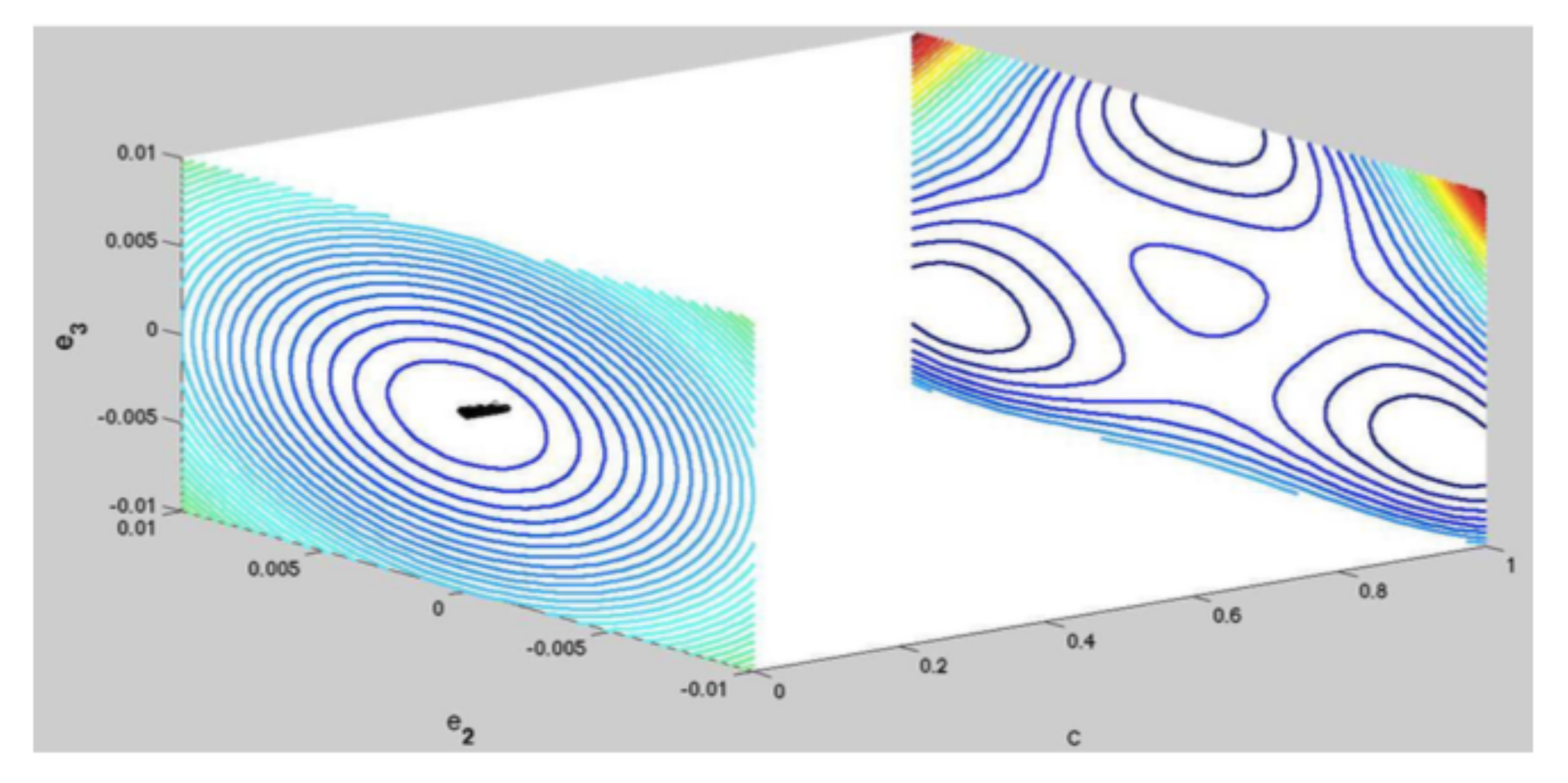}
\caption{\textbf{Movie S8} [MovieS8.mp4]: Evolution of material points on the homogeneous free energy density surface, $\mathscr{F}(c, \be)$ for the three-dimensional formulation, restricted to the $\{c,e_2,e_3\}$ sub-space. The microstructure that forms is similar to that in Movie S7.}
\end{figure}

\begin{figure}[hbt]
  \centering
  \includegraphics[width=0.5\textwidth]{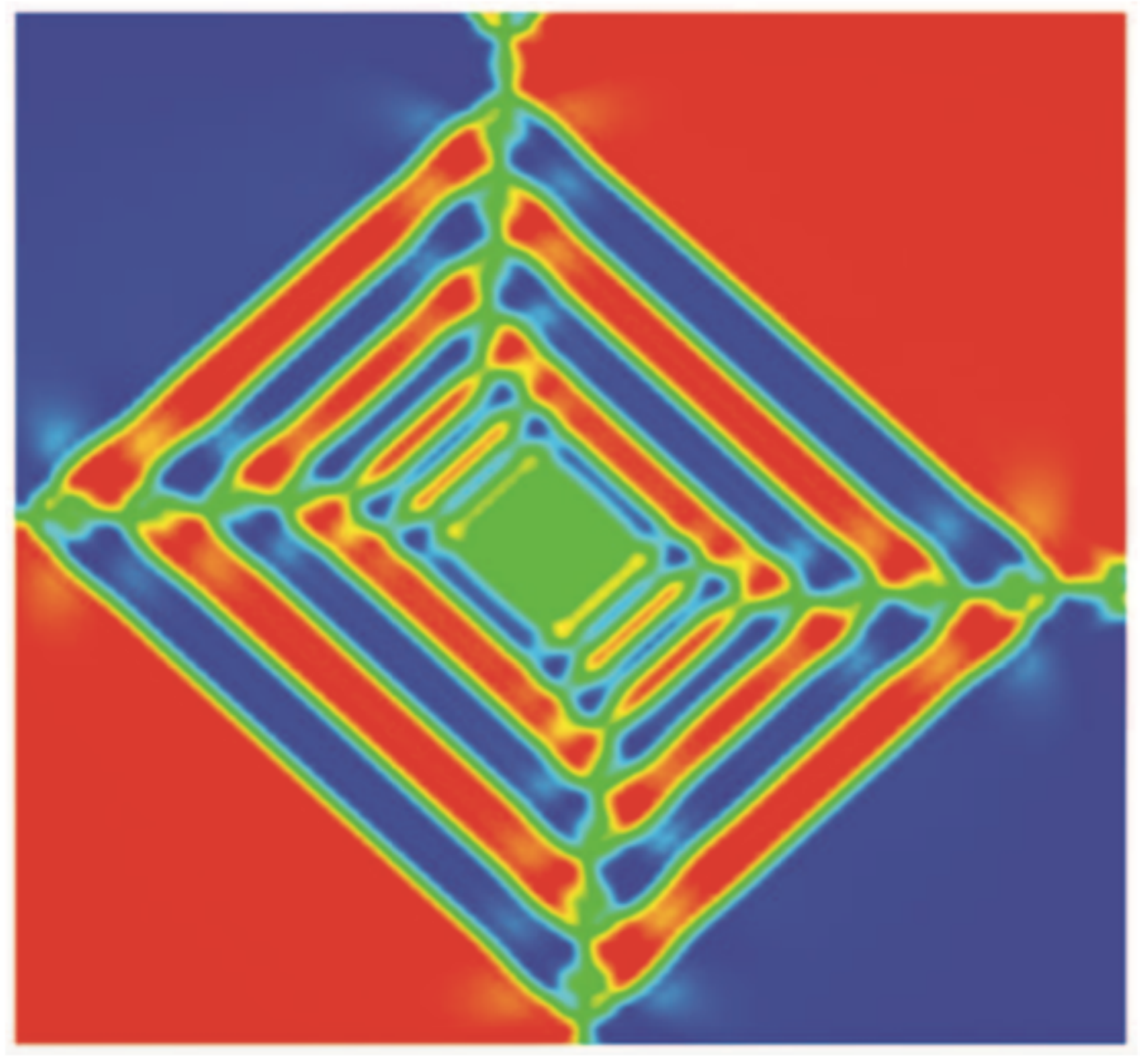}
\caption{\textbf{Movie S9} [MovieS9.mp4]: Evolution of a cross-hatched two-dimensional microstructure.}
\end{figure}

\begin{figure}[hbt]
  \centering
  \includegraphics[width=0.5\textwidth]{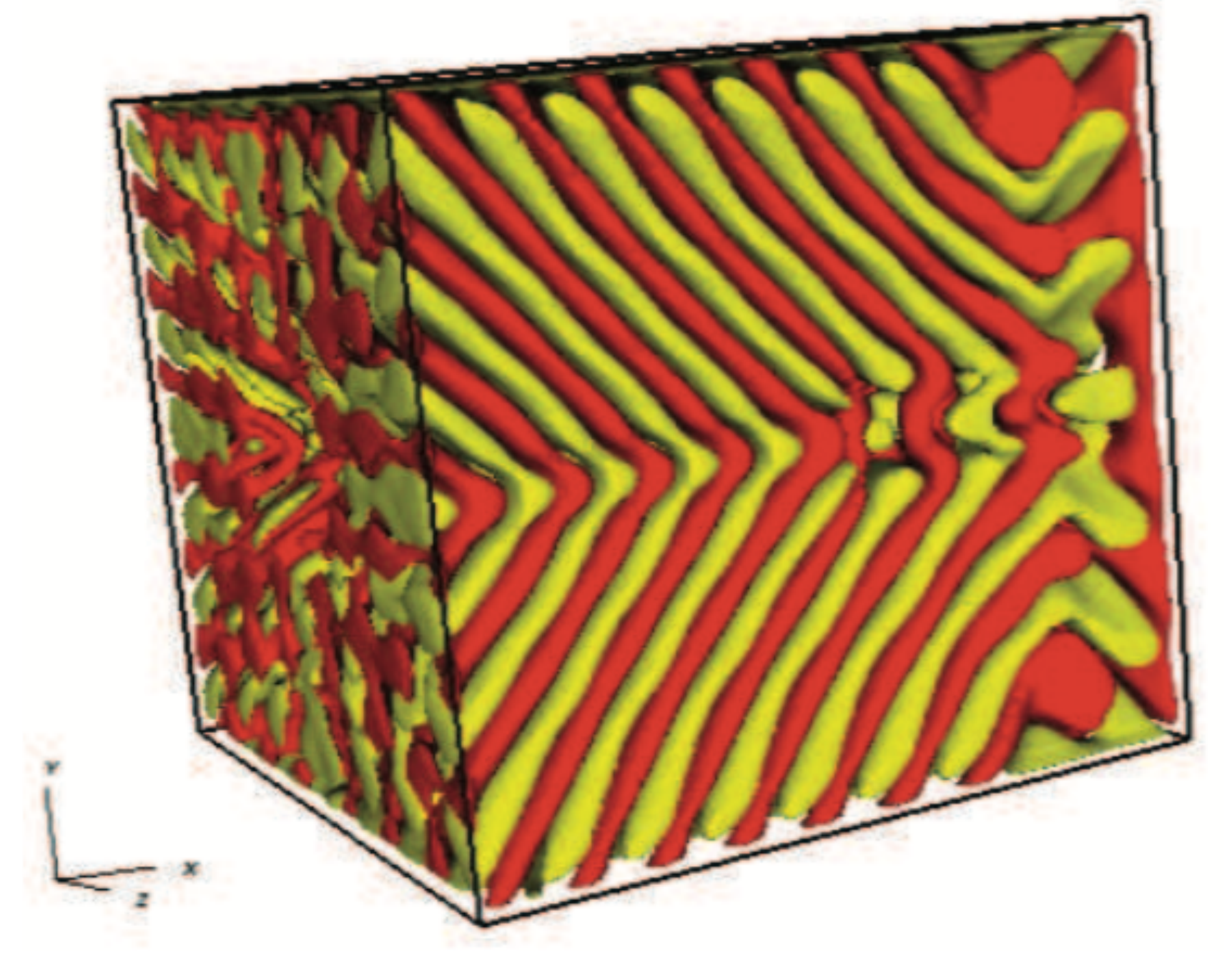}
\caption{\textbf{Movie S10} [MovieS10.mp4]: Evolution of plate-like structures in a three-dimensional microstructure.}
\end{figure}
